\documentclass[12pt]{article}
\usepackage[utf8]{inputenc}
\usepackage{graphicx}
\usepackage{amsmath,amssymb}
\usepackage[margin=1in]{geometry}
\usepackage{authblk}
\usepackage{setspace}
\usepackage{caption}
\usepackage{float}
\usepackage{physics}
\title{\bf Ge-based Quantum Sensors for Low-Energy Physics }

\author{D.-M. Mei, N. Budhathoki, S. A. Panamaldeniya, K.-M. Dong, S. Bhattarai, A. Warren, A. Prem, S. Chhetri}
\affil{University of South Dakota, 414 East Clark Street, Vermillion, South Dakota 57069, USA}
\date{}

\begin{document}

\maketitle

\begin{abstract}

We present \textbf{GeQuLEP} (Germanium-based Quantum Sensors for Low-Energy Physics), a conceptual design for an advanced quantum sensing platform integrating high-purity germanium (Ge) crystals with engineered phononic crystal cavities. At cryogenic temperatures, these cavities naturally host dipole-bound states, effectively forming quantum dots coupled to radio-frequency quantum point contact (RF-QPC) readout systems. This innovative coupling approach promises ultra-sensitive phonon-mediated charge detection through phonon-induced charge displacement. GeQuLEP is specifically designed to achieve exceptionally low detection thresholds, theoretically enabling single primary phonon sensitivity with anticipated energy depositions as low as \textbf{0.00745~eV}. This unprecedented sensitivity, if realized experimentally, would provide unique access to searches for low-mass dark matter down to the keV/$c^2$ mass range via nuclear and electronic recoils. Additionally, GeQuLEP aims to facilitate the real-time detection of solar \textit{pp} neutrinos through coherent elastic neutrino--nucleus scattering (CE$\nu$NS). By combining phonon-based quantum transduction with quantum-classical hybrid readout schemes, the GeQuLEP architecture represents a scalable, contact-free phonon spectroscopy design that could significantly advance the capabilities of ultra-low-energy rare-event detection at the quantum limit.

\end{abstract}

\section{Introduction}

Dark matter and neutrinos are two of the most enigmatic and foundational components of the universe. Together, they play critical roles in shaping the large-scale structure of the cosmos, the behavior of galaxies, and the thermodynamic evolution of the early universe. Dark matter, which constitutes approximately 85\% of all matter, remains undetected in laboratory experiments, yet its gravitational influence is well-established through astrophysical observations such as galactic rotation curves, gravitational lensing, and the cosmic microwave background~\cite{Bertone2005,Planck2018}. Unveiling the nature of dark matter—whether it consists of weakly interacting massive particles (WIMPs), light bosonic fields, or particles from hidden sectors—remains one of the foremost challenges in particle physics and cosmology.

Neutrinos, though extremely light and weakly interacting, are equally significant. They are the most abundant massive particles in the universe, playing a pivotal role in processes ranging from stellar fusion and supernovae to early universe dynamics and matter-antimatter asymmetry~\cite{Robertson2015,Formaggio2012}.Precise measurements of neutrino properties—such as the mass hierarchy, absolute mass scale, Majorana or Dirac nature, magnetic moment, and the existence of potential sterile states—could provide critical insights into physics beyond the Standard Model. Coherent elastic neutrino–nucleus scattering (CE$\nu$NS), in particular, provides a unique window into flavor-independent neutrino interactions, with applications in astrophysics, nuclear physics, and nonproliferation~\cite{Freedman1974,Akimov2021}.

Recent years have witnessed rapid progress in the experimental search for dark matter, with leading experiments such as XENONnT~\cite{Aprile2023}, LUX-ZEPLIN (LZ)~\cite{LZ2023}, and SuperCDMS~\cite{Agnese2018} pushing sensitivity to lower cross sections and masses. However, these experiments are fundamentally limited by their detection thresholds, which are typically in the range of 0.05--1~keV, leaving the MeV-scale dark matter parameter space~\cite{Essig2012,Battaglieri2017} largely unexplored. 

As described by Mei et al.~\cite{Mei2018}, low-mass dark matter particles interact too weakly to generate detectable signals in conventional detectors, depositing recoil energies well below their electronic noise floors. These energies, often below 1~eV, lie beyond the reach of charge collection and scintillation-based techniques that underpin today’s most sensitive platforms. As a result, an entire class of theoretically motivated dark matter candidates—those with MeV-scale masses or coupling through light mediators—remains effectively invisible to current searches.

A similar limitation arises in neutrino physics. Detecting low-energy neutrinos, such as solar \textit{pp} neutrinos, through CE$\nu$NS holds the potential to revolutionize solar and particle physics. CE$\nu$NS is a neutral-current process, offering a flavor-blind, enhanced cross-section for low-energy neutrinos. However, the associated nuclear recoil energies are extremely small—well below 1~eV—placing them out of reach for most current detectors. While the COHERENT experiment~\cite{Akimov2021} has measured CE$\nu$NS at higher neutrino energies using pulsed sources, extending this technique to continuous solar neutrino fluxes demands ultra-low-threshold instrumentation.

\begin{figure}
    \centering
    \includegraphics[width=0.45\textwidth]{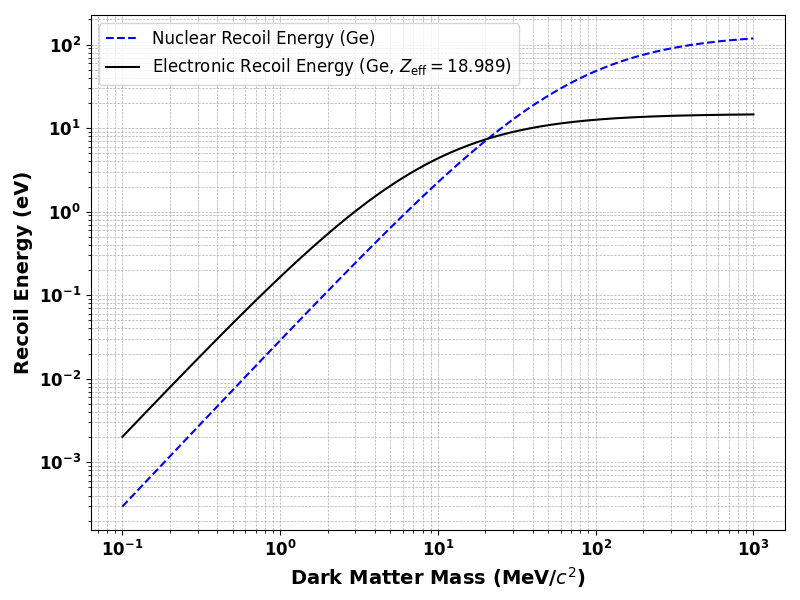}
    \includegraphics[width=0.45\textwidth]{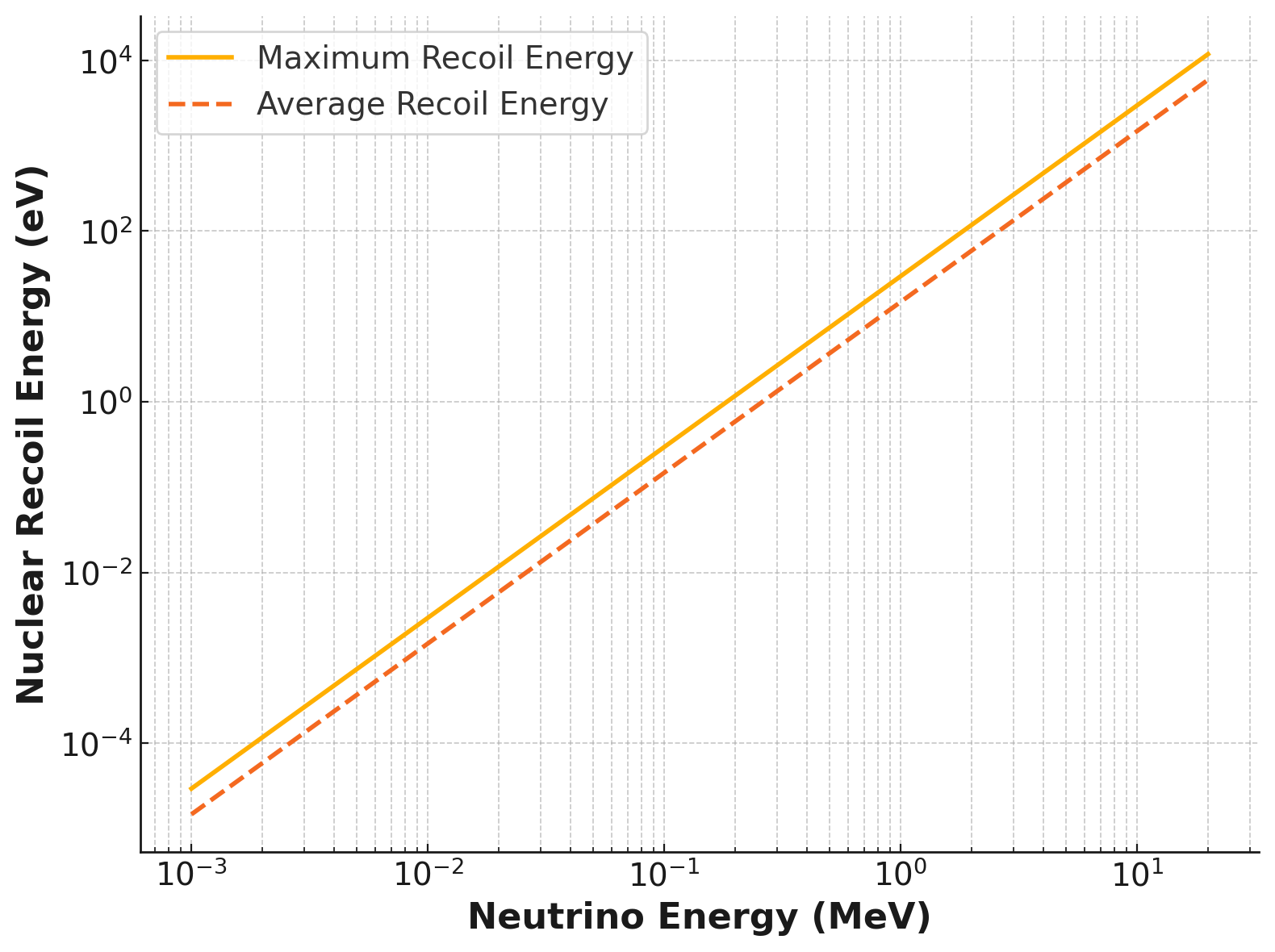}
    \caption{Left: Recoil energies in Ge as a function of dark matter mass, showing both nuclear and electronic channels~\cite{Mei2018}. Right: Predicted nuclear recoil spectrum in Ge from solar neutrinos via CEvNS interactions~\cite{Billard2014}.}
    \label{fig:recoil_spectra}
\end{figure}

Figure~\ref{fig:recoil_spectra} illustrates the challenge: For dark matter masses below $\sim$10~MeV/$c^2$, or for low-energy solar neutrinos, the energy depositions in a germanium (Ge) detector lie well below 1~eV. In particular, solar \textit{pp} neutrinos—dominant in the solar spectrum—generate nuclear recoils in the few-eV regime, while MeV-scale dark matter may produce even smaller electronic recoils. These signals are undetectable with existing technologies, limiting our ability to explore new physics in these low-energy domains.

These limitations collectively underscore the urgent need for next-generation detectors with energy thresholds reaching the $\sim$0.1~eV level or lower. Achieving such sensitivity is not merely a technical milestone—it is a gateway to probing entire sectors of particle physics that are currently inaccessible. From freeze-in dark matter production mechanisms to dark photon mediation, and from solar dynamics to neutrino magnetic moments, these physics questions demand a radically different approach to signal transduction and amplification.

While internal charge amplification in Ge has demonstrated the potential to achieve detection thresholds as low as 0.1~eV, further development of this technology is currently being pursued by Mei et al.~\cite{Mei2018}. Another promising approach is internal phonon amplification via the production of Neganov-Trofimov-Luke (NTL) phonons, as implemented by the SuperCDMS collaboration~\cite{Agnese2018}. This technique also shows potential for reaching sub-eV detection thresholds by amplifying phonon signals through the drift of charge carriers across the detector volume.

However, both existing detection technologies rely critically on drifting charges through the bulk material of the detector, necessitating complex fabrication processes, including highly reliable and ultra-low-noise electrical contacts. A fundamentally simpler approach—one that entirely avoids electrical contacts and eliminates charge transport through the bulk—would represent a substantial advancement. Ideally, such a detector would rely on direct phonon spectroscopy, sensitive to single primary phonons generated by minuscule energy depositions from dark matter or neutrino interactions. Currently, no established technology achieves this phonon-level sensitivity, representing a significant technological gap in the field of ultra-low-energy particle detection.

To bridge this technological gap, we introduce the Ge-based Quantum Sensors for Low-Energy Physics (GeQuLEP) detector platform. GeQuLEP pioneers a novel phonon spectroscopy approach, leveraging quantum wells (QWs)—analogous to gate-defined quantum dots—naturally formed from dipole-bound states in high-purity Ge crystals at cryogenic temperatures. These quantum dots localize single charge carriers while simultaneously trapping and slowing phonons within engineered quantum phononic crystal (PnC) cavities. By coupling these phonon-induced quantum interactions to ultra-sensitive radio-frequency quantum point contact (RF-QPC) readout systems, GeQuLEP enables direct, contact-free detection of individual phonon events, targeting unprecedented sensitivity thresholds of 0.01~eV or lower. This quantum-enabled phonon spectroscopy platform thus presents a highly scalable and inherently noise-resilient pathway for detecting exceptionally faint energy depositions, positioning GeQuLEP at the forefront of next-generation detectors for low-mass dark matter and low-energy neutrino interactions.

To systematically develop and evaluate the GeQuLEP phonon spectroscopy architecture, this paper is structured as follows: Section~\ref{sec:concept_design} presents the core concept and detailed design of the GeQuLEP platform, including the physics behind dipole-induced quantum dots, their integration into PnC cavities, and the principles underlying QPC-based readout. Section~\ref{sec:phonon_physics} explores the generation, propagation, and spectral characteristics of ballistic phonons in Ge at cryogenic temperatures, emphasizing phonon spectroscopy relevance. In Section~\ref{sec:surface_loss}, we analyze phonon reflectivity at the Ge–vacuum interface and identify potential loss mechanisms arising from surface imperfections and thin-film interactions. Section~\ref{sec:charge_coupling} quantitatively assesses phonon–charge coupling efficiency, describes charge displacement dynamics, and models the resultant signals detectable by the QPC readout. Section~\ref{sec:experiment_outlook} examines the experimental feasibility, highlighting fabrication challenges and proposing strategic implementation pathways. Finally, Section~\ref{sec:conclusion} summarizes our findings and underscores the broader implications of the GeQuLEP phonon spectroscopy platform for future searches for rare events in dark matter and neutrino physics.

\section{Concept and Design}
\label{sec:concept_design}
At cryogenic temperatures, quantum dots in Ge can be realized via two distinct mechanisms: (1) electrostatic confinement using gate-defined potentials and (2) impurity-induced localization via freeze-out of shallow dopants. Both mechanisms generate localized deformation potentials capable of confining single charge carriers and supporting quantized energy levels—ideal for low-energy quantum sensing applications.

In gate-defined quantum dots, voltages applied to lithographically patterned electrodes create local electric fields that modulate the band structure of the underlying semiconductor~\cite{Kouwenhoven1997, Hanson2007}. This results in a spatially varying deformation potential that confines charge carriers within a narrow region. The confinement potential is often modeled as a Gaussian~\cite{Burkard1999}:

\begin{equation}
V_{\text{gate}}(x) = -V_0 \exp\left(-\frac{x^2}{2\sigma^2}\right),
\end{equation}

where \(V_0\) is the potential depth and \(\sigma\) characterizes the spatial width. The deformation potential energy shift arises from electrostatic or strain fields and is described by~\cite{BirPikus1974, Mahan2000}:

\begin{equation}
\Delta E_{\text{DP}} = D \cdot \varepsilon(x),
\end{equation}

where \(D\) is the deformation potential constant (\(\sim 13.5\,\text{eV}\) for electrons in Ge~\cite{Jacoboni1983}) and \(\varepsilon(x)\) is the local strain or field gradient. The quantity \( \varepsilon(x) \) represents the local strain field and is defined as the relative change in length due to deformation. In one dimension, it is given by the spatial derivative of the displacement field, \( \varepsilon(x) = \partial u(x)/\partial x \), where \( u(x) \) is the local displacement at position \( x \)~\cite{LandauElasticity}. Strain is a dimensionless quantity, as it represents a normalized measure of deformation—specifically, the fractional change in length of a material element. In three dimensions, the strain field is described by the symmetric strain tensor \( \varepsilon_{ij} = \frac{1}{2}(\partial u_i/\partial x_j + \partial u_j/\partial x_i) \), and the scalar dilational strain relevant to deformation potential energy shifts is given by the trace \( \text{Tr}(\varepsilon) = \varepsilon_{xx} + \varepsilon_{yy} + \varepsilon_{zz} \)~\cite{Cleland2003}. This strain couples to charge carriers via the deformation potential, resulting in local shifts in the conduction or valence band edge energies.

A comparable, yet naturally occurring confinement mechanism emerges in high-purity Ge operated below 10~K. At these temperatures, shallow impurity atoms freeze out and form localized dipole states~\cite{Mei2024,Mei2022}, which create internal deformation potentials without the need for external gating. Specifically, boron, aluminum, and gallium acceptors give rise to p-type dipole states that bind holes, while phosphorus donors form n-type dipole states that bind electrons. These dipole states act as quantum dots, confining charge carriers through localized Coulombic potentials that can be modeled as:

\begin{equation}
V_{\text{dipole}}(x) = -\frac{q^2}{4\pi \varepsilon_0 \varepsilon_r d },
\end{equation}
which describes the electrostatic potential along the perpendicular bisector of an ideal electric dipole,  
where \(q\) is the elementary charge, \(d\) is the dipole size (i.e., the separation distance between positive and negative charges), \(\varepsilon_0\) is the vacuum permittivity, and \(\varepsilon_r \approx 16\) is the relative permittivity of Ge. 

The dipole size is constrained by the temperature at which the dipole states are thermally stabilized, and can be estimated by:
\begin{equation}
    d = \frac{q^2}{4\pi\epsilon_0\epsilon_r k_B T},
\end{equation}
where \( q \) is the elementary charge, \( \epsilon_0 \) is the vacuum permittivity, \( \epsilon_r \) is the relative permittivity of the material, \( k_B \) is the Boltzmann constant, and \( T \) is the temperature.

The Coulomb potential between a positive and a negative charge forming a dipole state is effectively constant, depending primarily on the fixed separation distance between the two charges. As shown in the equation above, this separation distance is determined by the thermal energy at which the dipole state is stabilized, and therefore varies only with temperature. Consequently, the spatial width of the resulting confinement potential is limited by the temperature-dependent variation of this dipole separation. Since temperature can be precisely controlled in cryogenic systems, the spatial extent of the confinement potential can also be tightly regulated, offering a tunable and stable platform for defining localized quantum states.

In practical modeling, especially at cryogenic temperatures in high-purity Ge, such dipole-induced potentials are often approximated by narrower, smooth confinement profiles—for example, Gaussian wells:

\begin{equation}
V_{\text{dipole}}(x) \approx -V'_0 \exp\left(-\frac{x^2}{2{\sigma'}^2}\right),
\end{equation}

where \(V'_0\) denotes the maximum potential depth (typically on the order of \(\sim\)10~meV), and \(\sigma'\) represents the characteristic confinement width (e.g., approximately 7~nm at 4~K, with possible variation depending on the variation of temperature). Both parameters can be extracted by fitting to experimental data~\cite{san, math}.

\begin{figure}[H]
    \centering
    \includegraphics[width=0.75\textwidth]{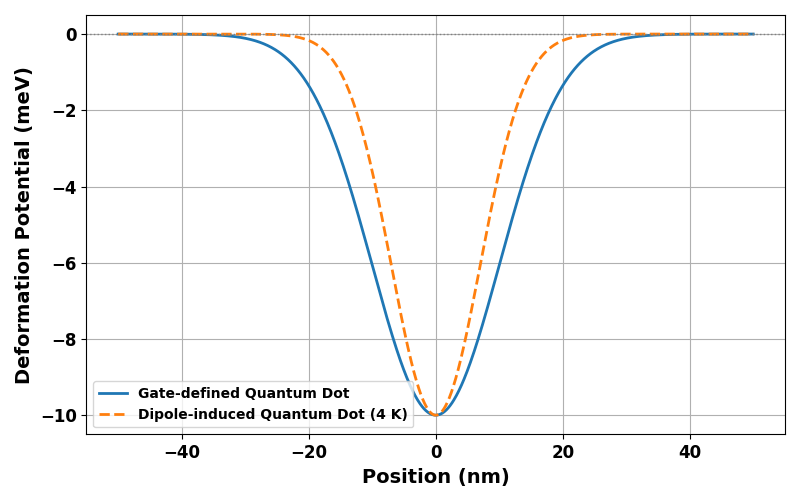}
    \caption{Comparison of deformation potentials in Ge: gate-defined quantum dot potential (solid blue) and dipole-induced quantum dot potential at 4~K (dashed orange). Dipole-induced potentials originate from frozen-out impurity atoms such as boron, aluminum, gallium, or phosphorus~\cite{Mei2024, Mei2022, san, math}.}
    \label{fig:deformation_comparison}
\end{figure}

The deformation potentials arising from gate-defined quantum dots and those naturally formed by impurity freeze-out at cryogenic temperatures exhibit remarkably similar spatial profiles, as illustrated in Figure~\ref{fig:deformation_comparison}. Both types of potentials are localized and can be effectively approximated by Gaussian wells, thereby creating confinement environments ideal for binding single charge carriers. The gate-defined potential (solid blue curve) is generated by external voltages that shape the semiconductor band structure through precise electrostatic control. In contrast, the dipole-induced potential (dashed orange curve) emerges intrinsically from localized electric fields produced by frozen-out impurities, such as boron, aluminum, gallium, or phosphorus. Despite their different physical origins, the comparable depths and widths of these potentials imply that impurity-induced dipole states closely mimic gate-defined quantum dots in their ability to confine carriers and support discrete energy levels. Consequently, these impurity-induced quantum dots present a naturally occurring and highly effective platform for quantum sensing applications in high-purity germanium.

In this work, we employ a Gaussian confinement potential as an approximation for the wavefunctions of dipole-bound quantum dots formed by shallow impurities in high-purity germanium. Although realistic impurity-induced potentials may differ significantly in symmetry and shape, the Gaussian model serves as a useful and widely adopted first-order approximation for analytically modeling the localization and spatial extent of hole states around impurity centers. This approach has been commonly utilized in the literature to simplify calculations and provide initial estimates of spin and charge interactions with phonons~\cite{Kloeffel2012, Maier2015}. Thus, our Gaussian approximation establishes a practical framework for determining baseline coupling strengths and evaluating the feasibility of the proposed quantum sensing architecture, forming a solid basis for future, more comprehensive numerical simulations or experimental investigations of impurity-bound states.

What makes this naturally occurring system particularly compelling is its seamless compatibility with established quantum readout techniques, such as QPCs. Just as gate-defined quantum dots are routinely integrated with nearby QPCs for ultra-sensitive charge detection in traditional semiconductor platforms, dipole-induced quantum dots in Ge can be similarly engineered to function as localized charge sensors. These impurity states—formed by controlled doping with elements like boron, aluminum, gallium, or phosphorus—appear at predictable spatial depths and can be reproducibly created with high precision. This allows them to be strategically positioned near the surface or edge of the detector, where QPCs can be patterned using standard lithographic techniques, enabling robust and scalable integration.

In addition to their exceptional stability and electrical tunability, impurity-bound quantum dots exhibit strong coupling to acoustic phonons via the deformation potential interaction. This property makes them highly suitable for phonon-mediated detection and positions them as promising elements for implementing phonon spectroscopy. When a ballistic phonon is absorbed, or a carrier is thermally excited or tunnels within a dipole potential, the resulting local charge redistribution can be sensitively detected by an adjacent QPC. When embedded in a RF reflectometry circuit, the QPC functions as a high-speed, high-sensitivity transducer, converting minute electrostatic fluctuations into measurable shifts in the amplitude and phase of the reflected RF signal. This architecture enables real-time, single-phonon sensitivity with sub-electron resolution and GHz-scale bandwidth, critical for detecting faint energy depositions from low-mass dark matter or neutrino interactions.

To further enhance phonon--charge coupling in the GeQuLEP architecture, the dipole-state layers ($\sim$ 1 $\mu$m) themselves can be engineered as embedded PnC cavities. By employing advanced nanoscale etching techniques, periodic arrays of holes or grooves are introduced directly into the doped regions, forming a two-dimensional acoustic lattice that generates phononic bandgaps—analogous to photonic crystals used to control light~\cite{Joannopoulos2008}. These bandgaps enable the selective trapping of phonons with target frequencies, confining them in the immediate vicinity of impurity-bound quantum states. The resulting localization significantly increases the interaction probability between phonons and dipole-bound carriers, thereby amplifying the induced signal. This integrated design allows the doped layer to serve a dual purpose: acting both as a quantum confinement region and as a phonon-resonant structure. It offers a compact, scalable, and contact-free solution for signal amplification while eliminating the need for external resonators and maintaining compatibility with quantum-limited RF-QPC readout systems.

In addition to enhancing phonon confinement, the ultrathin two-dimensional PnC layer also functions as an effective phonon filter that suppresses thermal background noise. By precisely tailoring the lattice geometry and periodicity, the structure can be tuned to create bandgaps that block low-energy phonons—typically those below 83~GHz at 4~K—associated with thermal excitations. This selective filtering prevents unwanted thermal phonons from reaching the QW region, thereby reducing spurious background signals. As a result, the signal-to-noise ratio for rare-event detection, such as interactions from low-mass dark matter and solar \(pp\) neutrinos, is substantially improved. The integration of this multifunctional PnC layer is thus pivotal for achieving high-fidelity phonon spectroscopy and robust signal discrimination in the GeQuLEP platform.

\begin{figure}[H]
    \centering
    \includegraphics[width=0.48\textwidth]{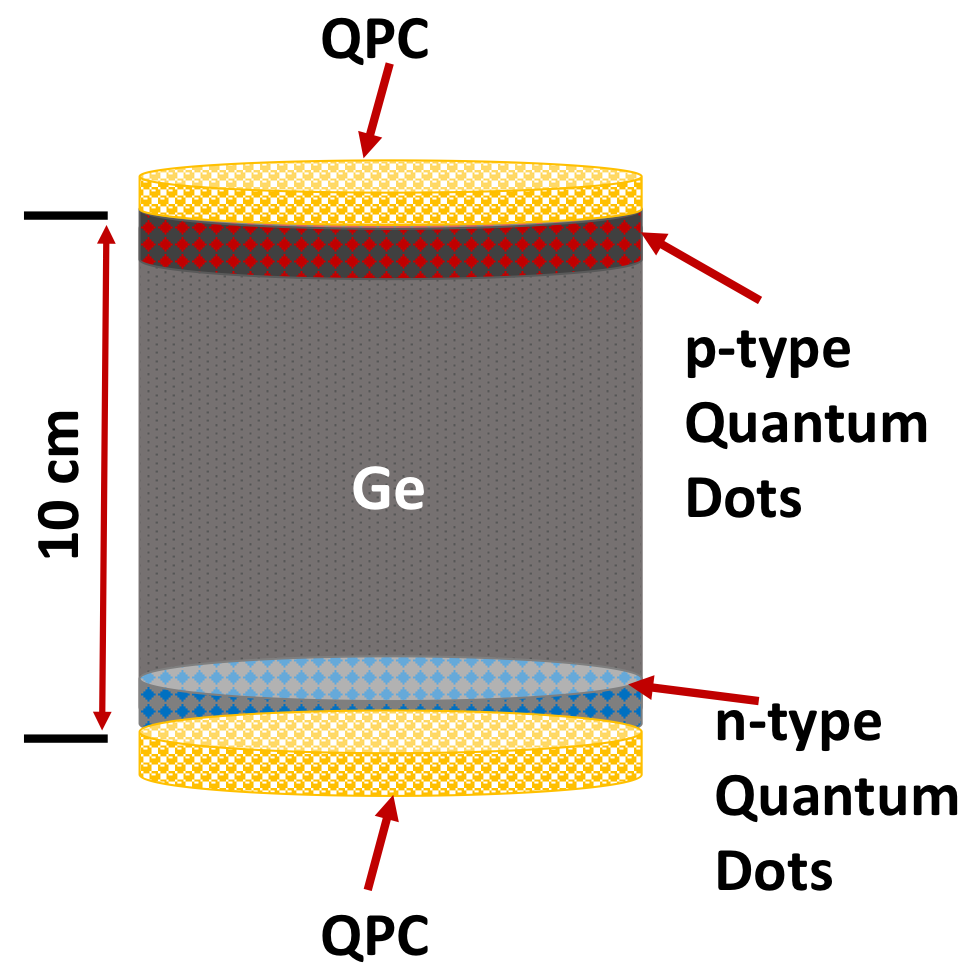}
    \includegraphics[width=0.48\textwidth]{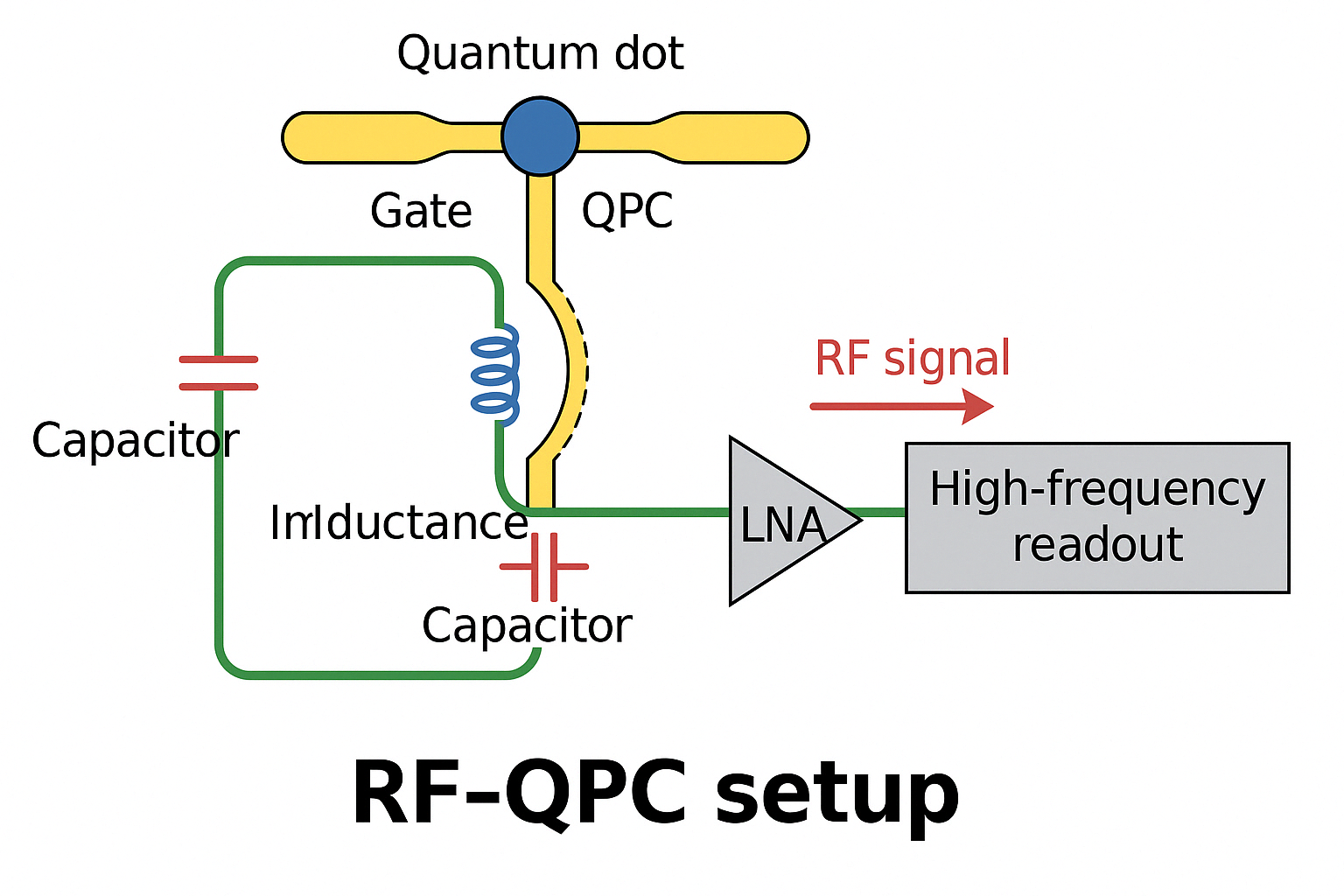}
    \caption{
    \textbf{Left:} Schematic of the GeQuLEP detector architecture. A high-purity germanium crystal with a central impurity concentration of approximately \(10^{10}~\text{cm}^{-3}\) forms the detector bulk. Near the top surface, p-type dipole quantum dots—naturally formed from boron, aluminum, or gallium dopants at concentrations of approximately \(\sim10^{14}~\text{cm}^{-3}\) in a single-crystal Ge substrate—are engineered into a PnC cavity structure. Likewise, n-type dipole quantum dots—formed from phosphorus dopants at similar concentrations—are embedded in a separate PnC cavity structure positioned near the bottom surface of the crystal. The PnC cavities with a thickness of 1 $\mu$m serve a dual purpose: trapping ballistic phonons within the quantum dot regions and filtering out thermal phonons to enhance signal fidelity. Adjacent QPCs are positioned to detect phonon-induced charge displacements. \textbf{Right:} RF-QPC readout configuration. Each QPC is embedded in an \(LC\) tank circuit designed to reflect an applied RF carrier signal. Phonon-induced motion of bound charges modulates the local electrostatic environment, thereby altering the QPC conductance. These charge variations are encoded onto the reflected RF signal and recovered through low-noise amplification (LNA) and demodulation electronics for signal analysis.
    }
    \label{fig:gequlep_readout}
\end{figure}

Figure~\ref{fig:gequlep_readout} illustrates the overall GeQuLEP detector architecture and its associated RF-QPC readout configuration. Spatially separated p-type and n-type dipole regions are strategically positioned at opposite ends of a high-purity Ge crystal to enable charge sensitivity across the full detector volume. PnC cavities surround these dipole regions, providing spatial and spectral confinement of ballistic phonons, while adjacent QPCs are precisely aligned to detect phonon-induced charge displacements. Together, these design elements enable the efficient transduction of low-energy phonon signals into measurable electronic responses, positioning the GeQuLEP platform as a scalable, ultra-sensitive quantum sensor ideally suited for phonon spectroscopy.

By eliminating the need for external gating and leveraging impurity engineering techniques already standard in semiconductor processing, this approach allows for scalable fabrication of quantum sensors that are inherently cryo-compatible and extremely low-noise. The combination of quantum confinement from dipole states, phonon-mediated interaction, and QPC-based charge readout constitutes the core innovation of the GeQuLEP platform. It offers a transformative pathway for detecting ultra-low-energy excitations such as those from MeV-scale dark matter and sub-eV solar neutrinos via CE$\nu$NS.

With the conceptual framework of GeQuLEP established—including the use of dipole-induced quantum dots, PnC confinement, and QPC-based readout—the next critical aspect is understanding the behavior of phonons that mediate signal transduction. In the following section, we examine how phonons are generated by low-energy particle interactions in Ge and how they propagate through the crystal lattice under cryogenic conditions, laying the foundation for their role in enabling sub-eV detection.

\section{Phonon Generation and Propagation in Ge}
\label{sec:phonon_physics} 

When a low-energy particle—such as a dark matter candidate or a neutrino—interacts with a Ge nucleus, it transfers a small amount of kinetic energy, typically in the sub-eV to sub-keV range. At these low energies, ionization in high-purity Ge is strongly suppressed due to its relatively large electronic bandgap of 0.73~eV at cryogenic temperatures (e.g., 4~K) and a high ionization threshold energy of approximately 2.96~eV per electron-hole pair. Consequently, the vast majority of the recoil energy is deposited into the crystal lattice and converted into quantized lattice vibrations, or \textit{phonons}~\cite{Luke1989,Shutt1992}. These energy depositions predominantly generate \textbf{primary phonons} in the form of longitudinal acoustic (LA) and transverse acoustic (TA) modes, with typical energies ranging from 1 to 10~meV~\cite{Shank1983,Haller1974}. Since the energy scale of these interactions lies well below the excitation threshold for optical phonons in Ge (30--40~meV), the generation of optical phonons is highly suppressed. Furthermore, the relatively small momentum transfer involved in such events favors the generation of long-wavelength acoustic modes. Among these, LA phonons dominate due to their stronger coupling to the deformation potential and higher group velocity, which enables efficient energy transport and interaction with localized charge carriers. These characteristics make LA and TA phonons—especially the former—the primary information carriers in phonon-mediated detection architectures such as GeQuLEP, which are optimized for capturing athermal acoustic phonons arising from rare, low-energy events like CE$\nu$NS and low-mass dark matter interactions.

To model phonon generation at a microscopic level, we consider a simplified one-dimensional lattice of atoms coupled by harmonic springs. The Lagrangian for this system of \( n \) coupled atoms of mass \( m \) connected by a spring constant \( k \) is given by
\begin{equation}
L = \sum_{i=1}^n \frac{1}{2}m\dot{x}_i^2 - \sum_{i=1}^n \frac{1}{2}k(x_i - x_{i-1})^2,
\end{equation}
where \( x_i \) denotes the displacement of the \( i \)-th atom from equilibrium. Applying the Euler-Lagrange equation yields the equation of motion:
\begin{equation}
m\ddot{x}_i = -k(x_i - x_{i-1}) + k(x_{i+1} - x_i).
\end{equation}
Assuming a traveling wave solution of the form \( x_i(t) = A e^{i(\omega t - qx_i)} \), where \( \omega \) is the angular frequency and \( q \) is the wavevector, we obtain the dispersion relation:
\begin{equation}
\omega^2 = \frac{2k}{m}(1 - \cos(qd)) \quad \Rightarrow \quad
\omega(q) = 2\sqrt{\frac{k}{m}} \sin\left(\frac{qd}{2}\right),
\end{equation}
where \( d \) is the lattice constant. The maximum vibrational frequency is attained at the Brillouin zone edge (\( q = \pi/d \)):
\begin{equation}
\omega_{\text{max}} = 2\sqrt{\frac{k}{m}}.
\end{equation}

To estimate the vibrational frequency scale for Ge, we use an atomic mass of $\approx 1.206 \times 10^{-25}~\text{kg}$,
and an effective spring constant \( k \approx 3.86~\text{N/m} \), based on its acoustic phonon modes~\cite{YuCardona2010}. Substituting these values, we obtain
$\omega_{\text{max}}  \approx 1.13 \times 10^{13}~\text{rad/s}$,
and the corresponding frequency:
\begin{equation}
f_{\text{max}} = \frac{\omega_{\text{max}}}{2\pi} \approx 1.80 \times 10^{12}~\text{Hz} = \boxed{1.80~\text{THz}}.
\end{equation}
This frequency corresponds to a maximum phonon energy of
$E = h f  \approx \boxed{7.45~\text{meV}}$,
which is consistent with the experimentally observed phonon energy spectrum in Ge~\cite{Agnese2018}.

The initial phonon distribution following a nuclear recoil is sharply peaked and non-thermal, with primary phonons generated throughout the Brillouin zone, including both zone-center and zone-edge modes. These high-frequency phonons quickly undergo \textbf{anharmonic decay} via three-phonon scattering processes such as:
\begin{equation}
\omega_0 \rightarrow \omega_1 + \omega_2,
\end{equation}
subject to conservation of energy and momentum:
\[
\omega_0 = \omega_1 + \omega_2, \quad \vec{q}_0 = \vec{q}_1 + \vec{q}_2.
\]
As the cascade continues, phonon energies are reduced until further scattering becomes inefficient. The resulting \textbf{ballistic phonons}, typically below 1~meV in energy, propagate quasi-freely over centimeter-scale distances at cryogenic temperatures (below 10~K)~\cite{Irwin1995,Haller1974}. These long-lived phonons retain the spatial and temporal information of the original recoil and are critical for signal transduction in ultra-sensitive phonon-based detectors.

\begin{figure} [H]
    \centering
    \includegraphics[width=0.48\textwidth]{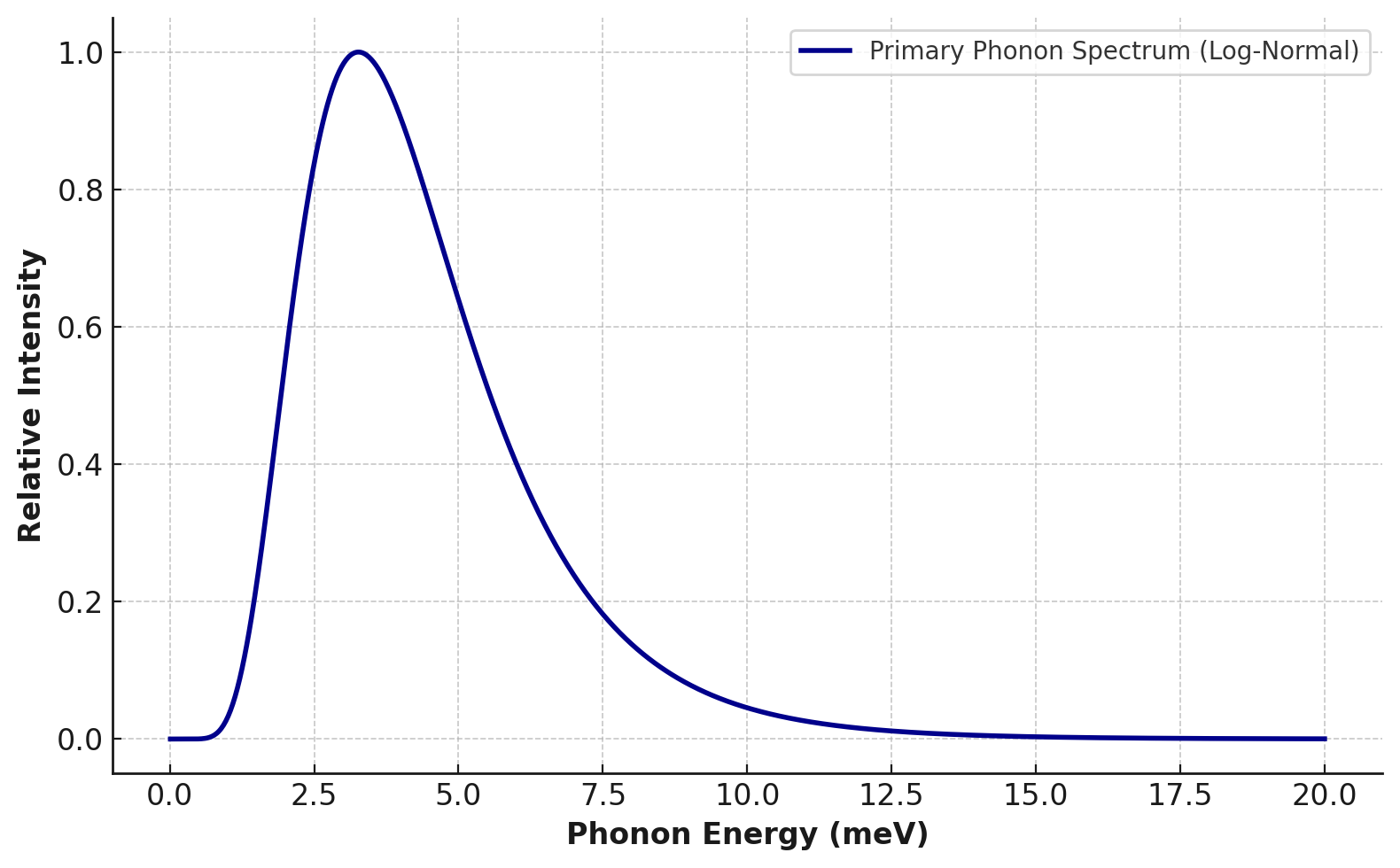}
    \includegraphics[width=0.48\textwidth]{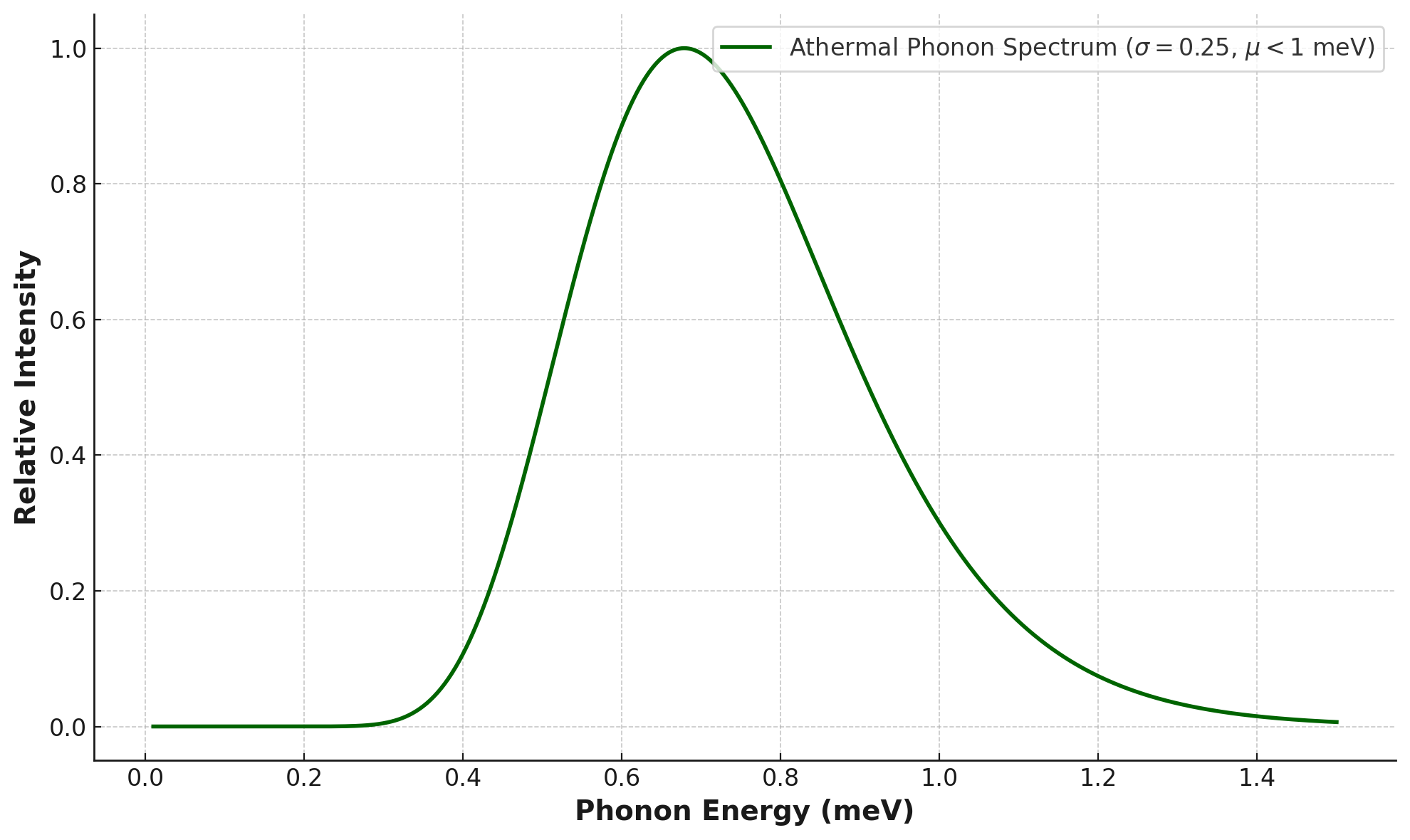}
    \caption{Left: Energy spectrum of primary phonons generated by nuclear recoil in Ge. Right: Energy spectrum of ballistic phonons after anharmonic downconversion, showing a broader distribution of lower-frequency phonons that propagate over long distances at cryogenic temperatures.}
    \label{fig:phonon_evolution}
\end{figure}

Figure~\ref{fig:phonon_evolution} illustrates this phonon evolution process in high-purity Ge—from the sharply peaked primary phonon spectrum to the broader, low-energy ballistic phonon distribution. These ballistic phonons are essential to the operation of cryogenic detectors such as GeQuLEP, where they can be absorbed by localized charge carriers in quantum dots or dipole states, and subsequently transduced into measurable electrical signals by QPCs.

Understanding the full pathway—from nuclear recoil to ballistic phonon propagation—is essential for optimizing the design of next-generation detectors targeting ultra-low-energy phenomena. In this context, \textit{phonon spectroscopy} emerges as a vital tool, enabling precise characterization of the phonon signatures generated by rare events. These capabilities are particularly critical for detecting MeV-scale dark matter and for observing CE$\nu$NS, where energy depositions occur well below the ionization threshold~\cite{Freedman1974, Billard2014, Agnese2018}.

For such low-energy events, a simplified one-dimensional model of phonon generation and transport provides a practical and justified approximation for estimating initial coupling strengths. These interactions typically deposit sub-eV to keV-scale energy in highly localized volumes, producing a small number of athermal phonons. In high-purity Ge at cryogenic temperatures, these phonons predominantly travel along preferred crystallographic directions due to strong anisotropic phonon focusing. Long mean free paths—especially for low-frequency LA phonons—allow energy to propagate coherently over centimeter-scale distances~\cite{Liu1999, Tamura1991, Perrin2006}. Given this behavior, one-dimensional modeling captures the dominant physical processes governing phonon-induced charge displacement, while also reducing computational complexity. This approximation thus serves as a robust starting point for phonon modeling in the GeQuLEP architecture, with comprehensive three-dimensional simulations for future work.

Equally important to phonon generation is the behavior of these ballistic phonons at material boundaries. In particular, the reflectivity of acoustic phonons at the Ge–vacuum interface plays a decisive role in determining phonon confinement, energy retention, and the overall signal yield. Understanding this interface interaction is critical for engineering phononic crystal cavities and guiding structures that enhance detection efficiency. The next section analyzes the effects of acoustic impedance mismatch at the Ge boundary and explores its implications for phonon trapping and waveguiding within the GeQuLEP detector system.

\section{Reflectivity of Acoustic Phonons at the Ge--Vacuum Interface}
\label{sec:surface_loss}

In high-purity Ge, acoustic phonons generated by nuclear recoil and subsequent anharmonic downconversion can propagate ballistically over macroscopic distances, particularly under cryogenic conditions where phonon-phonon and phonon-defect scattering are strongly suppressed. These low-frequency, phase-coherent excitations retain directional momentum, making them highly suitable for phonon-based sensing and signal transduction in quantum detectors.

The behavior of ballistic phonons at material interfaces critically influences their lifetime and detectability. For the GeQuLEP detector architecture, which exploits ballistic phonons as information carriers, understanding the reflectivity of the Ge--vacuum interface is essential. Owing to the extreme mismatch in specific acoustic impedance, this interface behaves as a nearly perfect acoustic reflector—analogous to Fresnel reflection in electromagnetic systems.

The specific acoustic impedance \( Z \) of a medium is given by
\begin{equation}
Z = \rho v,
\end{equation}
where \( \rho \) is the mass density and \( v \) is the speed of sound for the phonon mode of interest. For LA phonons—which dominate long-range ballistic transport—the specific acoustic impedance of Ge is approximately
$Z_{\text{Ge}} \approx 2.87 \times 10^7~\text{kg}/\text{m}^2\cdot\text{s}$,
based on \( \rho_{\text{Ge}} \approx 5323~\text{kg/m}^3 \) and \( v_{\text{Ge}} \approx 5400~\text{m/s} \). By contrast, vacuum cannot support acoustic wave propagation, yielding \( Z_{\text{vac}} = 0 \).

The reflection coefficient \( R \) for a longitudinal wave normally incident on an interface between two media with specific acoustic impedances \( Z_1 \) and \( Z_2 \) is given by
\begin{equation}
R = \left( \frac{Z_2 - Z_1}{Z_2 + Z_1} \right)^2.
\end{equation}
At the Ge--vacuum boundary, this becomes
\begin{equation}
R = \left( \frac{0 - Z_{\text{Ge}}}{0 + Z_{\text{Ge}}} \right)^2 = 1,
\end{equation}
indicating total reflection of incident acoustic phonons. This result is consistent with empirical data from cryogenic phonon detectors, which report negligible energy loss across optically polished Ge surfaces~\cite{Mazin2002}.

For oblique incidence, Snell’s law for acoustic waves predicts a critical angle
\begin{equation}
 \theta_c = \arcsin(v_{\text{vac}} / v_{\text{Ge}}) = 0^\circ , 
\end{equation}
confirming that total internal reflection occurs at all angles. Consequently, ballistic phonons remain confined within the Ge crystal, undergoing specular or quasi-specular reflections depending on surface quality and phonon wavelength.

This high reflectivity is foundational to the GeQuLEP detector’s operation. By confining phonons within the detector volume and minimizing inelastic losses, the architecture enables extended phonon path lengths, increased interaction probabilities, and enhanced energy resolution for sub-eV event detection.

While the idealized Ge--vacuum interface exhibits near-perfect reflectivity, practical deviations arise due to finite geometries and real surface conditions. In particular, thin-film structures and non-ideal surfaces introduce interference phenomena and phonon scattering that can reduce coherence and detection fidelity.

In thin Ge layers, multiple reflections between opposing surfaces lead to standing wave patterns analogous to Fabry–Pérot resonances in optics. Constructive interference occurs when the condition
\begin{equation}
2t \cos \theta = m \lambda, \quad m \in \mathbb{Z^+},
\end{equation}
is satisfied, where \( t \) is the film thickness, \( \lambda \) is the phonon wavelength, \( \theta \) is the internal angle of propagation, and the symbol \( \mathbb{Z}^+ \) refers to the set of positive integers. Under such resonant conditions, phonon transmission and energy distribution become frequency dependent~\cite{SwartzPohl1989}, leading to modulations in reflectivity and phonon lifetime. These effects can introduce temporal dispersion and degrade energy resolution in detectors relying on precise phonon timing, such as GeQuLEP.

Even in bulk systems, surface imperfections introduce additional phonon loss channels that compromise ballistic transport. Major mechanisms include:
\begin{itemize}
  \item \textbf{Surface roughness}, which causes diffuse scattering and disrupts phase coherence, reducing signal strength at the sensor interface~\cite{Klitsner1988}.
  \item \textbf{Near-surface crystal defects and dislocations}, which promote inelastic scattering and thermalization of athermal phonons~\cite{Haller1974}.
  \item \textbf{Contamination and oxidation}, which modify the acoustic impedance at the interface, enabling energy leakage into amorphous or lossy modes.
\end{itemize}

Mitigating these effects is essential for maximizing the performance of phonon-mediated quantum sensors. Techniques such as atomic-scale surface polishing, in situ passivation, and controlled interface engineering have demonstrated efficacy in preserving phonon coherence. For the GeQuLEP platform, which depends on the detection of individual ballistic phonons to achieve sub-eV energy thresholds, surface optimization is a key design requirement for maintaining signal integrity and maximizing quantum efficiency.

The high reflectivity of the Ge--vacuum interface, coupled with suppressed inelastic scattering at cryogenic temperatures, ensures efficient confinement and prolonged coherence of ballistic acoustic phonons within the detector volume. This physical advantage underpins the feasibility of using phonons as robust carriers of quantum information and energy deposition signatures. However, to fully harness their potential in sensing architectures like \text{GeQuLEP}, it is essential to evaluate how these confined phonons interact with embedded quantum structures. In the following section, we investigate the \textit{phonon detection efficiency} and explore the mechanisms of \textit{charge--phonon coupling} in QWs hosting dipole-defined quantum dots, which serve as the active transduction sites in the phonon-based quantum sensing framework and form the core of a novel phonon spectroscopy approach for detecting ultra-low-energy events.

\section{Phonon Detection and Charge–Phonon Coupling in Dipole-Defined Quantum Dots}
\label{sec:charge_coupling}

The detection of ballistic phonons in Ge-based quantum sensors relies critically on their efficient coupling to charge carriers confined in localized quantum dots. In the GeQuLEP detector platform, such quantum dots are naturally formed from dipole states at cryogenic temperatures and interact with incident phonons primarily through deformation potential coupling. The overall phonon detection efficiency depends on three key factors: (1) the geometric collection efficiency of phonons reaching the QW, (2) the phonon survival probability during propagation, and (3) the phonon-to-charge conversion efficiency, governed by the charge--phonon coupling strength. We analyze each of these contributions below to evaluate the resulting detection efficiency and the induced charge signal at the QPC.

\subsection{Geometric Collection Efficiency}

Primary acoustic phonons generated by low-energy particle interactions in Ge are emitted quasi-isotropically from the interaction site. A naive estimate of the fraction reaching the QW is given by the subtended solid angle:
\begin{equation}
\eta_{\text{geom}} = \frac{\Omega_{\text{QW}}}{4\pi}.
\end{equation}

However, this simple model neglects a crucial effect: \textbf{total internal reflection} at the Ge--vacuum interface. Owing to the extreme mismatch in acoustic impedance, phonons striking the boundary at any normal and non-normal angle are almost entirely reflected back into the bulk, as discussed in Section~\ref{sec:surface_loss}. This near-unity reflectivity causes phonons to remain confined within the Ge crystal, undergoing multiple specular or quasi-specular reflections.

As a result, the probability that a phonon eventually reaches the QW region increases significantly over time. This confinement effect can be further enhanced by incorporating PnC structures that guide and focus phonon propagation toward the sensing region. In optimized geometries, the effective collection efficiency can approach unity. Accordingly, we adopt a higher value for \( \eta_{\text{geom}} \) in our analysis to reflect minimal phonon escape and enhanced guiding toward the quantum dot region.

\subsection{Phonon Survival in Bulk Ge and Absorption in the Quantum Well}

In the GeQuLEP detector architecture, only phonons that successfully reach the QW region—where the QPC is located—can contribute to a measurable signal through charge–phonon coupling~\cite{Zhang2021, Cleland2020}. Therefore, understanding the \textit{phonon survival probability}, i.e., the likelihood that a phonon remains unabsorbed during its propagation from the point of generation to the QW region, is critical for evaluating detector performance~\cite{SwartzPohl1989, Klitsner1988}.

The primary loss mechanism during propagation is phonon absorption by unintended coupling to dipole states distributed throughout the Ge crystal. Phonons that interact with these off-path dipole states induce charge displacements away from the QPC sensing region, effectively dissipating their energy and quantum information before they can be detected. This parasitic absorption is governed by the spatial density of dipole states and the strength of the charge–phonon coupling~\cite{Feldman1993, Phipps2016}.

Assuming a uniform dipole density \( n_d \) and an average phonon absorption cross section \( \sigma_{\text{abs}} \), the probability that a phonon is absorbed over a path length \( L \) can be described using the Beer–Lambert law:
\begin{equation}
P_{\text{abs}} = 1 - \exp(-n_d \sigma_{\text{abs}} L).
\end{equation}
Accordingly, the survival probability becomes:
\begin{equation}
\eta_{\text{surv}} = \exp(-n_d \sigma_{\text{abs}} L).
\label{eq:absorption}
\end{equation}

To estimate \( \sigma_{\text{abs}} \), we apply Fermi’s Golden Rule, which provides the transition rate \( \Gamma \) for a phonon being absorbed by a two-level dipole state~\cite{Ziman1972, Mahan2000}:
\[
\Gamma = \frac{2\pi}{\hbar} |M_{if}|^2 \rho(\omega),
\]
where \( |M_{if}|^2 \) is the phonon–dipole interaction matrix element, and \( \rho(\omega) \) is the phonon density of states. For LA phonons interacting via the deformation potential mechanism, the matrix element is approximated by~\cite{Mahan2000}:
\[
|M_{if}|^2 \approx D^2 \cdot \frac{\pi v_{\text{ph}}}{\rho V \omega^2},
\]
where \( D \) is the deformation potential constant, \( \rho \) is the mass density of the medium, and \( V \) is the normalization volume. The phonon density of states in 3D is \( \rho(\omega) = \omega^2 / (2\pi^2 v_{\text{ph}}^3) \), where \( v_{\text{ph}} \) is the phonon group velocity.

Dividing \( \Gamma \) by the incident phonon flux \( \Phi = v_{\text{ph}} / V \) gives the absorption cross section:
\[
\sigma_{\text{abs}} = \frac{\Gamma}{\Phi} = \Gamma \cdot \frac{V}{v_{\text{ph}}},
\]
which simplifies to:
\begin{equation}
   \sigma_{\text{abs}} = \frac{D^2}{\rho v_{\text{ph}}^3 \hbar}.
   \label{eq:crosssection}
\end{equation}

Using this relation and adopting deformation potential values for Ge at 4~K—approximately 13.5~eV for electrons and 4.5~eV for holes~\cite{Mahan2000, Cleland2020}—we find absorption cross sections of \( \sigma_{\text{abs}} \approx 5.3 \times 10^{-13}~\text{cm}^2 \) for electrons and \( \sim 5.9 \times 10^{-14}~\text{cm}^2 \) for holes.
 These values align well with literature reports on impurity-mediated phonon absorption in Ge at cryogenic temperatures~\cite{SwartzPohl1989, Klitsner1988, Phipps2016}, validating the model. Equation~\ref{eq:absorption} thus encapsulates the cumulative survival probability for a ballistic phonon, emphasizing the importance of minimizing unintentional dipole densities in the bulk to ensure high QPC sensitivity.

\begin{figure}[H]
    \centering
    \includegraphics[width=0.7\linewidth]{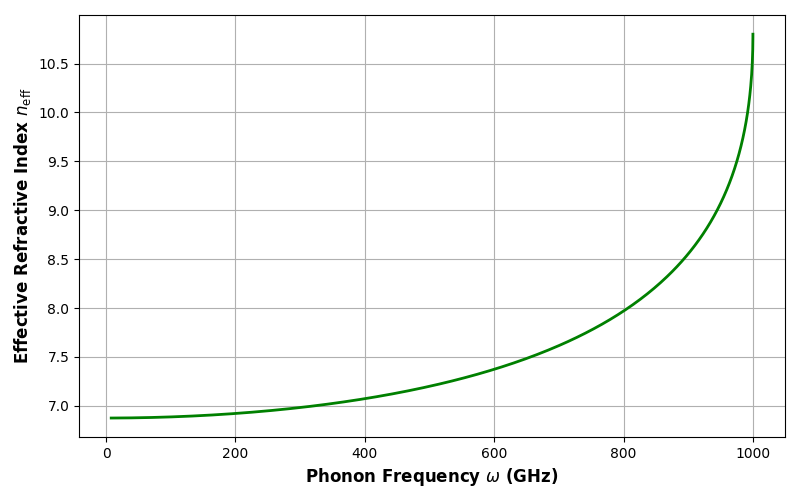}
    \caption{Effective refractive index $n_{\mathrm{eff}}$ as a function of phonon frequency $\omega$ based on a synthetic dispersion relation in a SiGe PnC cavity.}
    \label{fig:neff}
\end{figure}

To quantify phonon propagation in the engineered QW region, we introduce an \textit{effective phonon refractive index}, \( n_{\mathrm{eff}}^{\mathrm{phonon}} \), inspired by optical analogies~\cite{joannopoulos2008}. In a structured PnC cavity, the phonon dispersion is modified, causing phonons to slow down substantially near band edges. We adopt a synthetic dispersion relation:
\[
\omega(q) = 2 \cdot \sin\left( \frac{\pi}{2} \cdot \frac{q}{q_{\mathrm{max}}} \right),
\]
where \( \omega \) is the frequency and \( q \) is the wavevector. The corresponding phase velocity is \( v_{\mathrm{ph}}(q) = \omega(q)/q \), and the effective refractive index is defined as:
\[
n_{\mathrm{eff}}(q) = \frac{v_{\mathrm{bulk}}}{v_{\mathrm{ph}}(q)} = \frac{v_{\mathrm{bulk}} \cdot q}{\omega(q)},
\]
with \( v_{\mathrm{bulk}} \approx 5.4~\text{mm}/\mu\text{s} \) for LA phonons in Ge. Figure~\ref{fig:neff} shows that \( n_{\mathrm{eff}} \) increases substantially in the low-frequency limit, where band flattening enhances phonon–matter interactions~\cite{Maldovan2013, mohammadi2009, safavi2019}.

Consequently, the phonon absorption cross section increases in the QW region, where the phase velocity of phonons are significantly slowed by the elevated effective refractive index, as shown in Figure~\ref{fig:neff}. This phonon slowing enhances confinement and strengthens interactions with localized dipole states that act as quantum dots. As predicted by Eq.~\ref{eq:crosssection}, the absorption cross section \( \sigma_{\text{abs}} \) is inversely proportional to the cube of the phase velocity. 

\begin{figure}[H]
    \centering
    \includegraphics[width=0.7\linewidth]{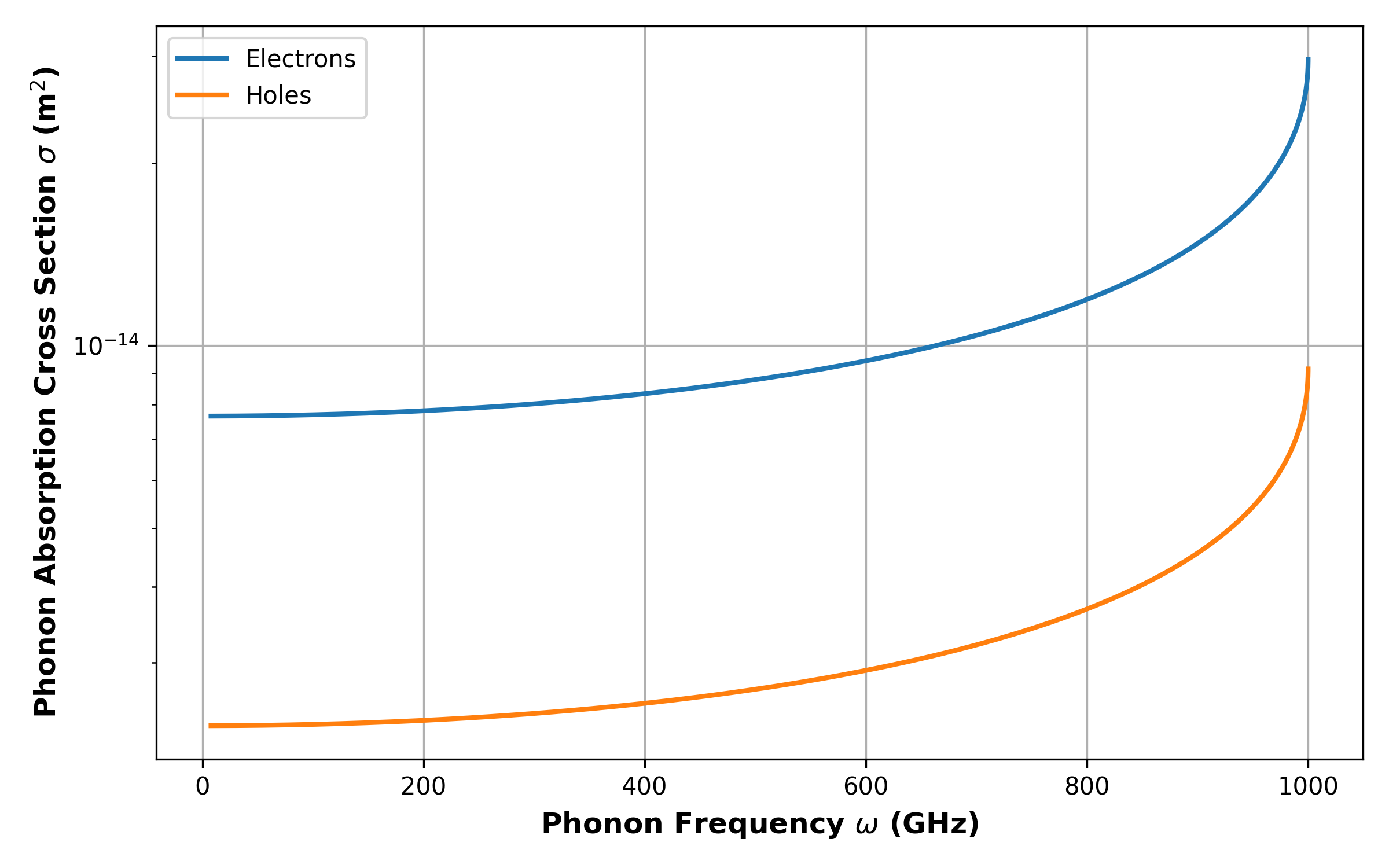}
    \caption{Phonon absorption cross section $\sigma$ as a function of phonon frequency in the quantum well region. The enhanced absorption at higher frequencies corresponds to the regime where phonons are strongly confined and slowed, increasing the likelihood of interaction with dipole states.}
    \label{fig:sigma_qw}
\end{figure}

Figure~\ref{fig:sigma_qw} illustrates that \( \sigma_{\text{abs}} \) increases sharply at high frequencies in the PnC-enhanced QW region, where phonons are strongly confined. These phonons are more likely to be absorbed by dipole states, inducing localized charge displacement detectable via QPC readout. Such enhanced phonon absorption is essential for realizing single-phonon resolution in next-generation quantum sensing platforms.

\subsection{Charge–Phonon Coupling Strength and Displacement}

The interaction between lattice vibrations (phonons) and charge carriers bound in dipole states is mediated by the deformation potential coupling mechanism. In this process, phonons dynamically modulate the local strain field, thereby altering the potential landscape experienced by electrons or holes confined in quantum dots formed by dipole states. This modulation induces small displacements of the bound charges, effectively converting phonon energy into an electrical response. The efficiency of this transduction pathway is determined by the strength of the deformation potential coupling, which governs the detector’s sensitivity to individual phonon events and its overall ability to register ultra-low-energy excitations.

The deformation potential coupling strength for an acoustic phonon mode quantifies the interaction between phonons and charge carriers bound in dipole states. It is given by~\cite{YuCardona2010}:
\begin{equation}
g(\omega) = D \cdot \frac{\omega}{v_{ph}} \cdot \sqrt{\frac{\hbar}{2 \rho V_{mode} \omega}} = D \cdot \sqrt{\frac{\omega \hbar}{2 \rho V_{mode} v_{ph}^2}},
\label{eq:couplingstrangth}
\end{equation}
where \( D \) is the deformation potential constant for Ge—approximately \( 13.5~\text{eV} \) for electrons and \( 4.5~\text{eV} \) for holes—\( \omega = 2\pi f \) is the phonon angular frequency, \( v_{\text{ph}} \approx 5400~\mathrm{m/s} / n_{\text{eff}} \) denotes the phase velocity of LA phonons in the QW region, \( \rho = 5323~\mathrm{kg/m}^3 \) is the mass density of Ge, and \( V_{\text{mode}} \) represents the phonon mode volume.

The mode volume \( V_{\text{mode}} \) characterizes the spatial region over which phonon energy is localized within a PnC cavity. Analogous to photonic crystal resonators, this volume can be estimated using the effective refractive index \( n_{\text{eff}} \), which captures the reduction in group velocity caused by the periodicity and band engineering of the PnC~\cite{Laude2015,Khelif2016,Maldovan2013}. For a phonon with wavelength \( \lambda \), the mode volume scales as
\[
V_{\text{mode}} \sim \left( \frac{\lambda}{n_{\text{eff}}} \right)^3,
\]
where \( \lambda = v_{\text{ph}} / f \), with \( f \) being the phonon frequency~\cite{joannopoulos2008}.

As an illustrative example, at \( f = 30~\mathrm{GHz} \), the corresponding phonon wavelength is \( \lambda \approx 180~\mathrm{nm} \) in Ge. Assuming an effective refractive index \( n_{\text{eff}} = 6.8 \)—a value consistent with high-index contrast PnC structures operating in the slow-phonon regime~\cite{Mohammadi2009,Hatanaka2014,Arrangoiz2016}—the estimated mode volume is \( V_{\text{mode}} \sim (180~\mathrm{nm}/6.8)^3 \approx 1.85\times10^{-5}~\mu\mathrm{m}^3 \), in good agreement with experimentally measured values in high-Q phononic resonators.

\begin{figure} [H]
    \centering
    \includegraphics[width=0.75\textwidth]{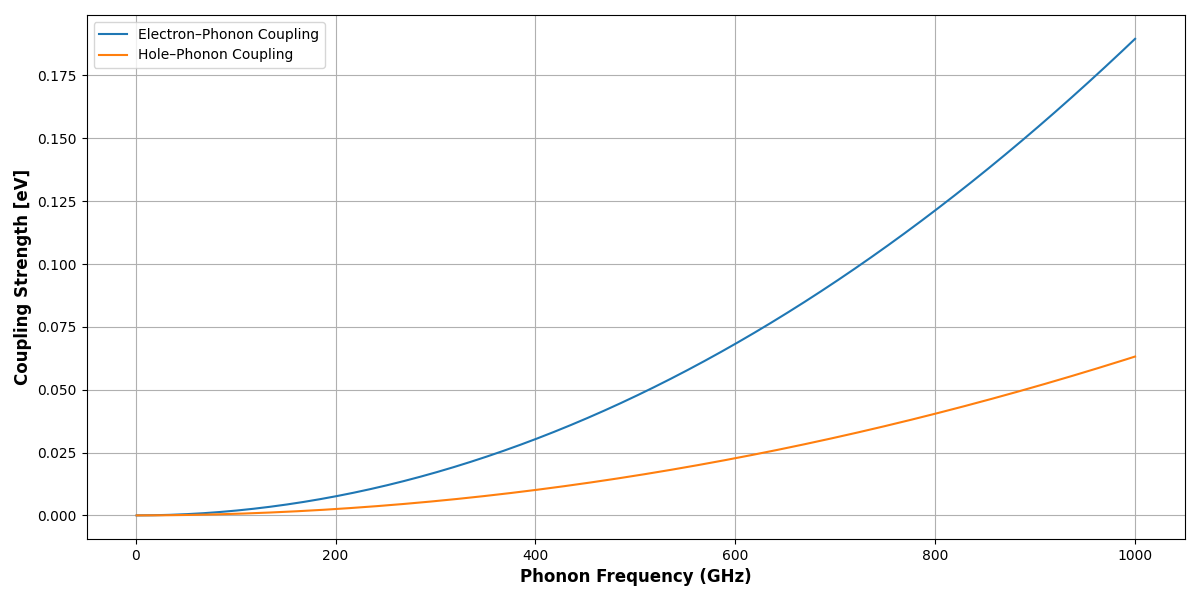}
    \caption{Charge–phonon coupling strength \( g(\omega) \) in Ge as a function of phonon frequency. The coupling increases with frequency due to the increase in phonon momentum.}
    \label{fig:phonon_coupling}
\end{figure}

As evident from Eq.~\ref{eq:couplingstrangth}, the coupling strength scales as \( g(\omega) \propto \sqrt{\omega} \), indicating that higher-frequency phonons interact more strongly with carriers via the deformation potential. Figure~\ref{fig:phonon_coupling} shows the calculated coupling strength over a frequency range from 1~GHz to 1~THz, illustrating the increasing interaction strength with frequency.

\subsubsection{Lattice Displacement Induced by a Single Phonon}

The lattice displacement associated with a single phonon is fundamentally linked to its coupling strength and wavelength. For an acoustic phonon of angular frequency \( \omega \), the wavelength is given by \( \lambda = 2\pi v_{ph} / \omega \), where \( v_{ph} \) is the phase speed of phonons in the medium~\cite{Kittel2005, AshcroftMermin}. Within the deformation potential framework~\cite{Mahan2000}, the charge--phonon coupling strength \( g(\omega) \) is defined as:
\begin{equation}
g(\omega) = D \cdot \frac{x_{\text{lattice}}}{\lambda} = D \cdot \frac{x_{\text{lattice}} \cdot \omega}{2\pi v_{ph}},
\end{equation}
where \( D \) is the deformation potential constant and \( x_{\text{lattice}} \) represents the characteristic lattice displacement induced by the phonon.

From quantum mechanical principles, the displacement amplitude associated with a single phonon mode confined within a mode volume \( V_{\text{mode}} \) in a crystal of mass density \( \rho \) is given by~\cite{Mahan2000, Ziman2001}:
\begin{equation}
x_{\text{lattice}}(\omega) = \pi \sqrt{\frac{2\hbar}{\omega \rho V_{\text{mode}}}}.
\end{equation}

\begin{figure}
    \centering
    \includegraphics[width=0.75\textwidth]{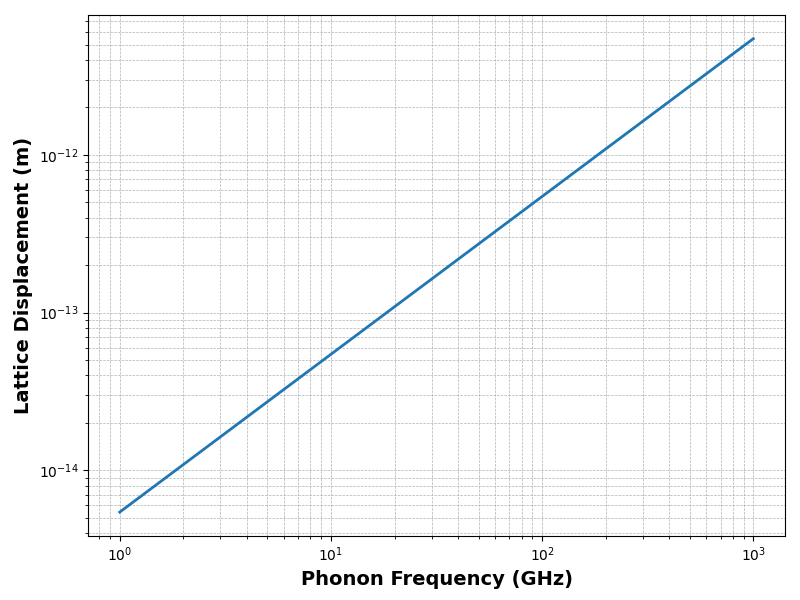}
    \caption{Lattice displacement \( x_{\text{lattice}} \) as a function of phonon frequency in Ge. Higher-frequency phonons result in smaller displacements due to shorter wavelengths.}
    \label{fig:lattice_displacement}
\end{figure}

This expression reveals that the lattice displacement scales inversely with the square root of the phonon frequency: \( x_{\text{lattice}} \propto \omega^{-1/2} \). Physically, this means that higher-frequency phonons produce smaller atomic displacements due to their shorter wavelengths and higher energies. Figure~\ref{fig:lattice_displacement} illustrates the calculated \( x_{\text{lattice}} \) across the GHz to THz frequency range for Ge.

As shown in Figure~\ref{fig:lattice_displacement}, the lattice displacement amplitude associated with a single phonon ranges from approximately \(10^{-14}\) to \(10^{-12}\) m, depending on the phonon frequency. These values correspond to extremely subtle perturbations of the atomic lattice, characteristic of quantum-scale vibrational excitations.
Because the ionic lattice charge and the bound charge in a dipole state have opposite signs, their respective contributions to the induced charge at the QPC can partially cancel. As a result, such minute lattice displacements typically yield a very weak induced signal at the QPC. In contrast, larger charge displacements—resulting from stronger charge-phonon interactions can produce significantly higher induced charges, effectively dominating the QPC response. Therefore, accurately quantifying phonon-induced charge displacement is essential for evaluating detection thresholds and optimizing the sensitivity of QPC-based phonon sensors.

\subsubsection{Charge-Phonon Induced Displacement of Bound Carriers}

At cryogenic temperatures such as \( 4~\mathrm{K} \), charge carriers in high-purity Ge or in the QW region can become thermally localized at impurity sites, forming dipole states. These localized carriers experience a restoring Coulomb force from the ionized impurity, resulting in bound states that behave effectively as quantum harmonic oscillators. Understanding the dynamics of these dipole states—particularly their interaction with athermal phonons—is essential for the development of phonon-coupled detectors and quantum sensors.

The electrostatic interaction between a bound charge and its host impurity is governed by the Coulomb potential:
\begin{equation}
    U(r) = -\frac{e^2}{4\pi \epsilon_0 \epsilon_r r},
\end{equation}
where \( e \) is the elementary charge, \( \epsilon_0 \) is the vacuum permittivity, \( \epsilon_r \) is the relative dielectric constant of Ge, and \( r \) is the distance between the charge and the ionic impurity.

To approximate the behavior of this bound system near its equilibrium position \( r_0 \), we expand the potential \( U(r) \) in a Taylor series around \( r_0 \). Neglecting the linear term due to equilibrium force cancellation, we obtain:
\begin{equation}
    U(r) \approx U(r_0) + \frac{1}{2}k(r - r_0)^2,
\end{equation}
where the effective spring constant \( k \) is given by the second derivative of the potential:
\begin{equation}
    k = \left| \frac{d^2 U}{dr^2} \right|_{r_0} = \frac{e^2}{2\pi \epsilon_0 \epsilon_r r_0^3}.
\end{equation}

The equilibrium radius \( r_0 \), also known as the \textit{Onsager radius}, is determined by balancing the Coulomb potential energy with the thermal energy of the carrier~\cite{Onsager1938}.

\begin{equation}
    \frac{e^2}{4\pi \epsilon_0 \epsilon_r r_0} = k_B T,
\end{equation}
where \( k_B \) is Boltzmann’s constant and \( T \) is the temperature. At \( T = 4~\mathrm{K} \), this relation yields an equilibrium radius of \( r_0 \approx 257.43~\mathrm{nm} \), which defines the thermal localization length of the bound charge. Using this value of \( r_0 \), the corresponding spring constant for the dipole state is calculated to be \( k \approx 1.67 \times 10^{-9}~\mathrm{N/m} \).

When an athermal phonon interacts with such a bound charge, it exerts an oscillatory force that perturbs the equilibrium, leading to an additional displacement. The dynamics of this perturbation are described by the driven harmonic oscillator equation:
\begin{equation}
    m \ddot{x} + kx = F_\text{phonon}(t),
\end{equation}
where \( m \) is the effective mass of the carrier and \( F_\text{phonon}(t) = F_0 \cos(\omega t) \) represents the external phonon-induced driving force. The resulting motion provides the foundation for charge sensing mechanisms in phonon-coupled readout systems such as QPCs.

The steady-state solution for displacement is:
\begin{equation}
    \delta x(t) = \frac{F_0/m}{|\omega_0^2 - \omega^2|} \cos(\omega t - \phi)
\end{equation}
where $\omega_0 = \sqrt{k/m}$ is the natural frequency of the bound particle.

The force exerted by a phonon on a charge carrier arises from the deformation potential interaction and is directly related to the spatial variation of the phonon displacement field \( u(x,t) \). Specifically, for a plane-wave phonon mode, the phonon-induced force can be expressed as:
\begin{equation}
    F_0 = D q^2 u_0,
\end{equation}
where \( D \) is the deformation potential constant, \( q \) is the phonon wavevector, and \( u_0 \) denotes the phonon displacement amplitude. 

The amplitude \( u_0 \) is estimated from the quantized displacement field corresponding to a single acoustic phonon mode in a crystalline solid. For a mode with angular frequency \( \omega \), this amplitude is given by~\cite{Mahan2000, Kittel2005}:
\begin{equation}
    u_0 \sim \sqrt{ \frac{ \hbar }{ 2 \rho V_{\text{mode}} \omega } },
\end{equation}
where \( \rho \) is the mass density of the crystal and \( V_{\text{mode}} \) is the effective mode volume of the phonon.

Using representative parameters for high-purity Ge at cryogenic temperatures—namely, a mass density \( \rho \approx 5.3 \times 10^3~\mathrm{kg/m^3} \), a phonon angular frequency \( \omega \sim 10^{11}~\mathrm{rad/s} \), and a mode volume \( V_{\text{mode}} \sim 0.000313~\mu\mathrm{m}^3 \)—the phonon displacement amplitude is estimated to be \( u_0 \sim 1.8 \times 10^{-14}~\mathrm{m} \). This value is several orders of magnitude below the interatomic spacing, consistent with expectations for single-phonon excitations in crystalline semiconductors at \( T \sim 4~\mathrm{K} \), where quantum zero-point motion dominates over thermal vibrations~\cite{YuCardona2010}.

The resulting displacement amplitude \( A \) of a bound charge, modeled as a driven harmonic oscillator, is then given by:
\begin{equation}
    \boxed{A = \frac{D q^2 u_0 / m}{|\omega_0^2 - \omega^2|}},
\end{equation}
where \( m \) is the effective mass of the carrier, and \( \omega_0 \) is the natural oscillation frequency of the bound state. This expression highlights the resonant enhancement of charge displacement when the phonon frequency approaches the natural frequency \( \omega_0 \), a regime critical to achieving high phonon-to-charge transduction efficiency.

\begin{figure} [H]
    \centering
    \includegraphics[width=0.75\textwidth]{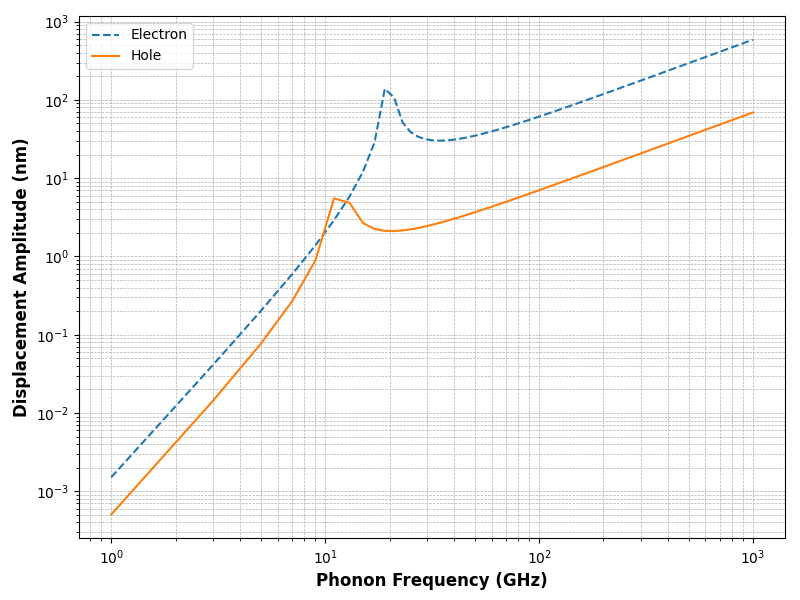}
    \caption{Calculated charge displacement \( x_{\text{charge}} \) as a function of phonon frequency for electrons and holes. A resonance enhancement is observed between 10--30~GHz, where the phonon frequency matches the natural oscillation of the dipole states. Beyond this range, displacement increases slightly with increasing frequency.}
    \label{fig:charge_displacement}
\end{figure}

Figure~\ref{fig:charge_displacement} illustrates the calculated charge displacement as a function of phonon frequency for both electrons and holes in Ge, using deformation potential constants of \( D_e = 13.5~\text{eV} \) for electrons and \( D_h = 4.5~\text{eV} \) for holes~\cite{YuCardona2010}. The results reveal a pronounced resonance region in the frequency range of approximately 10--30~GHz, where the phonon frequency closely matches the natural oscillation frequency of bound dipole states. Within this regime, the phonon-induced force is strongly amplified, resulting in a significant enhancement in charge displacement.

However, the resonant frequency band in the 10--30\,GHz range can also be thermally populated at cryogenic temperatures (e.g., 4\,K), leading to unwanted phonon-induced background noise. Thus, while this frequency range is optimal for enhancing phonon-to-charge conversion efficiency due to resonance with quantum dot states, it is concurrently susceptible to thermal contamination. This dual nature presents both a challenge and an opportunity: effective suppression of thermal background is crucial for preserving signal integrity, yet the resonant enhancement can be strategically harnessed to amplify the transduction of signal phonons arising from rare low-energy deposition events when necessary.

At frequencies above 30\,GHz, charge displacement continues to increase gradually with rising phonon frequency, consistent with the shorter wavelengths and more localized lattice deformation of high-energy phonons. Importantly, electrons consistently exhibit greater displacement amplitudes than holes across the full spectrum, suggesting a performance advantage for targeting electron-bound quantum dots in Ge. This insight underscores the potential of electron-based quantum dots for achieving higher sensitivity in phonon-mediated quantum sensing and spectroscopy applications.

As illustrated in Figure~\ref{fig:charge_displacement}, phonon-induced charge displacements in the resonant frequency range of 10--30\,GHz can exceed 1\,nm. Even at frequencies above 100\,GHz, the charge displacement remains significant, on the order of \(10^{-9}\) m. This displacement is nearly four orders of magnitude greater than the corresponding lattice displacement, highlighting a pronounced mechanical-to-electrical transduction at the charge sensor interface. Such substantial displacements are well within the detection sensitivity limits of advanced RF-QPCs~\cite{Schoelkopf1998, GonzalezZalba2015}. The exceptional charge sensitivity offered by RF-QPC technology positions it ideally for phonon-mediated signal readout in rare-event detection applications.

In our phonon-induced charge displacement model, we utilize a simplified harmonic oscillator approximation to represent the coupling between phonon modes and localized carrier states. This approach assumes an isotropic harmonic confinement potential and does not account for complexities such as crystal anisotropy, variations in impurity distributions, and more intricate impurity-host interactions inherent to realistic Ge quantum dots. While this simplified model provides essential initial insights and practical estimates of coupling strengths, we acknowledge that actual quantum dot environments, especially those resulting from impurity-induced potentials, may significantly deviate from these ideal conditions. Future work will incorporate detailed numerical simulations and targeted experimental investigations to capture these complexities more accurately, thereby refining our understanding of phonon-carrier interactions in high-purity germanium quantum dot structures.

This analysis demonstrates that individual phonons generated in high-purity Ge can produce detectable charge displacement in the quantum well region, affirming the feasibility of phonon-based quantum transduction within the GeQuLEP detector architecture. The robust deformation potential coupling, combined with the sensor's sensitivity to sub-eV phonon energies, underscores the promise of GeQuLEP as a powerful platform for detecting rare, low-energy depositions associated with low-mass dark matter interactions and solar neutrinos.

While the enhancement of charge displacement is critical, it represents just one component of the detector’s overall sensitivity. Achieving a measurable signal also depends on how efficiently ballistic phonons are collected and guided to the sensing region, and how effectively their energy is transduced into an induced charge. The following subsection explores these considerations by analyzing the phonon collection efficiency and the resulting electrostatic signal at the QPC.

\subsection{Phonon Collection Efficiency and Induced Charge}

The efficiency of phonon-based signal transduction in the GeQuLEP detector depends on two primary factors: the ability to guide phonons from their point of origin to the sensing region, and their survival during propagation. This combined effect is best described as the \textit{phonon collection efficiency}, defined as:
\begin{equation}
\eta_{\text{collection}} = \eta_{\text{geom}} \cdot \eta_{\text{prop}},
\end{equation}
where: 
\begin{itemize}
    \item \( \eta_{\text{geom}} \) is the geometric collection factor, accounting for the fraction of phonons that are directed toward the sensing region. With the aid of PnC cavity structuring and near-unity reflectivity at the Ge--vacuum interface, this factor can approach \( \eta_{\text{geom}} \sim 1.0 \), as discussed in Section~\ref{sec:surface_loss}.
    
    \item \( \eta_{\text{prop}} \) is the phonon propagation survival probability, representing the fraction of ballistic phonons that avoid inelastic scattering or absorption during transit. For a large path length \( L = 10~\text{cm} \), the survival probability is given by:
    \[
    \eta_{\text{prop}} = \exp(-n_d \sigma_{\text{abs}} L) \approx 0.95,
    \]
    where \( n_d = 10^{10}~\text{cm}^{-3} \) is the density of dipole states in a typical high-purity Ge crystal, \( \sigma_{\text{abs}} \approx 5.3\times 10^{-13}~\text{cm}^2 \) is the phonon absorption cross-section as evaluated using Eq.~\ref{eq:crosssection}, and \( L = 10~\text{cm} \) is the average distance phonons travel to reach the QW region~\cite{SwartzPohl1989, Klitsner1988}.
\end{itemize}

Substituting representative values:
\[
\eta_{\text{collection}} \approx 1.0 \times 0.95 \approx 0.95.
\]
\vspace{1em}
\noindent
This analysis shows that even for a large GeQuLEP detector, the probability that ballistic phonons reach the QW region and interact effectively with charge carriers confined in dipole-bound states remains exceptionally high—approximately 95\% for thick detectors and approaching 100\% for devices with a thickness of 1~cm. The phonon absorption probability in the QW region can be estimated using the relation
\[
P_{\text{abs}} = 1 - \exp(-n_d \sigma_{\text{abs}} L),
\]
where \( n_d \) is the dipole state density, \( \sigma_{\text{abs}} \) is the phonon–carrier absorption cross section, and \( L \) is the effective interaction length within the QW region. For typical parameters, such as \( n_d = 10^{14}~\text{cm}^{-3} \), \( \sigma_{\text{abs}} = 10^{-10}~\text{cm}^2 \), and \( L = 1~\mu\text{m} \), the calculated absorption probability \( P_{\text{abs}} \) approaches 100\%, indicating highly efficient phonon capture. This high efficiency is enhanced by PnC cavities near the band edges, which slow down phonons and increase their interaction cross section, as illustrated in Figure~\ref{fig:sigma_qw}. Consequently, nearly all of the 16 ballistic phonons produced through anharmonic decay from a primary phonon are expected to be absorbed and transduced via the nearby QPC electrodes.

\subsection{Phonon Down-Conversion through Anharmonic Decay}

At temperatures below 10\,K, the dominant scattering mechanism for high-frequency phonons in crystalline solids is three-phonon anharmonic decay. In this process, a single high-energy phonon decays into two lower-energy phonons, facilitated by the intrinsic anharmonicity of the lattice potential. For LA phonons, the decay rate exhibits a strong frequency dependence and is described by~\cite{SwartzPohl1989, Klitsner1988, Msall1997}:
\begin{equation}
\tau_{\text{ph}}^{-1} = 1.61\times10^{-55}~\text{s}^4 \cdot \nu^5,
\end{equation}
where \( \nu \) is the phonon frequency in Hz. For primary phonons with frequencies close to 1.8\,THz - typical of those generated by energy depositions from nuclear recoils or particle interactions in Ge - the corresponding lifetime is approximately \( \tau_{\text{ph}} \sim 0.33~\mu\text{s} \). This results in a characteristic diffusion length of roughly 0.6~$\mu$m before the phonon undergoes its first anharmonic decay.

The anharmonic decay cascade continues until the phonons reach sufficiently low energies, typically around 0.0005\,eV (corresponding to a frequency of approximately 125\,GHz) at which point inelastic scattering is strongly suppressed and the phonons transition into the ballistic regime. Each stage of the decay process produces two daughter phonons with roughly half the energy of the parent phonon, meaning more than four sequential decay events are required to reduce a primary 1.8\,THz phonon to the ballistic regime. As a result, a single high-energy primary phonon can yield up to \( 2^4 = 16 \) ballistic phonons.

These ballistic phonons retain both phase coherence and directional momentum, allowing them to traverse macroscopic distances with minimal scattering. Depending on the geometric collection efficiency and internal boundary reflections within the detector, a significant fraction—up to approximately 95\%—can reach the QW region, where the QPC is located. This analysis suggests that a single energy deposition of 0.00745\,eV can generate roughly 16 ($2^4$) ballistic phonons capable of coupling to localized charge carriers. These interactions are expected to produce measurable charge displacements, thereby enabling ultra-sensitive detection within the GeQuLEP architecture.

As these ballistic phonons approach the sensing region, they encounter dipole states embedded within a PnC cavity specifically engineered to filter, trap, and concentrate phonon flux. This cavity structure increases the likelihood of charge–phonon interactions by extending phonon dwell times and guiding phonon trajectories toward localized charge centers. A considerable fraction of these phonons are thus funneled into interactions with charge carriers confined in dipole-defined quantum dots. These interactions predominantly occur via deformation potential scattering, wherein phonon-induced distortions in the dipole structure modulate the local electrostatic potential experienced by the charge carriers. The resulting perturbations lead to measurable charge signals in the adjacent QPC.

The magnitude and detectability of these phonon-induced signals depend critically on the spatial and temporal distribution of phonons, their energy, and the confinement characteristics of the dipole states. In the following subsection, we analyze how these phonon-induced charge displacements are transduced into measurable signals at the QPC, forming the central mechanism behind the phonon spectroscopy capabilities of the GeQuLEP quantum sensing platform.

\subsection{Induced Charge on the QPC}

To quantify the signal generated by a phonon-induced displacement of a bound charge in a quantum dot, we apply the Ramo–Shockley theorem, which is widely used in the analysis of charge induction in solid-state detectors~\cite{Ramo1939, Spieler2005}.

The Ramo–Shockley theorem states that the instantaneous current \( I(t) \) induced on an electrode by a moving point charge \( Q \) is given by:
\begin{equation}
    I(t) = Q \cdot \mathbf{v}(t) \cdot \nabla \phi_w,
\end{equation}
where \( \mathbf{v}(t) \) is the velocity of the charge and \( \phi_w \) is the weighting potential. The weighting potential is defined by solving Laplace's equation,
\[
\nabla^2 \phi_w = 0,
\]
with boundary conditions such that \( \phi_w = 1 \) on the electrode of interest (e.g., the QPC) and \( \phi_w = 0 \) on all other surrounding electrodes.

To compute the total induced charge \( Q_{\text{ind}} \) on the QPC due to a displacement \( \delta x \) of the bound charge, we integrate the induced current over time:
\begin{equation}
    Q_{\text{ind}} = \int I(t)\,dt = Q \cdot \Delta \phi_w,
\end{equation}
where \( \Delta \phi_w \) is the change in weighting potential caused by the displacement:
\begin{equation}
    \Delta \phi_w = \phi_w(x + \delta x) - \phi_w(x) \approx \delta x \cdot \left. \frac{d \phi_w}{dx} \right|_{x}.
\end{equation}
Assuming a planar geometry where the QPC is modeled as a parallel electrode at a distance \( d \) from the charge, the weighting potential varies linearly with position: \( \phi_w(x) = x/d \). This gives
\[
\left. \frac{d \phi_w}{dx} \right|_{x} = \frac{1}{d}.
\]
Substituting into the expression for \( Q_{\text{ind}} \), we obtain:
\begin{equation}
    \boxed{Q_{\text{ind}} = Q \cdot \frac{\delta x}{d}},
\end{equation}
which shows that the induced charge is linearly proportional to the phonon-induced displacement \( \delta x \), the elementary charge \( Q \), and inversely proportional to the distance \( d \) between the quantum dot and the QPC electrode.
This result emphasizes the sensitivity of QPC-based detection schemes to nanoscale displacements, making them ideal for readout of phonon-coupled quantum systems.

\begin{figure} [H]
    \centering
    \includegraphics[width=0.75\textwidth]{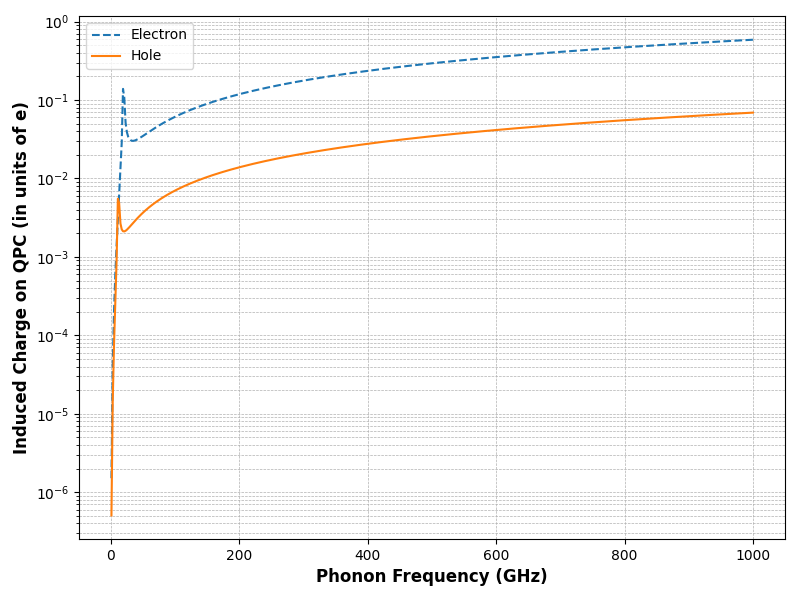}
    \caption{Induced charge \( Q_{\text{ind}} \) on the QPC electrode as a function of phonon frequency for electrons and holes in Ge, assuming a quantum dot–QPC separation of 1~\(\mu\)m. A pronounced resonance enhancement occurs in the 10--30~GHz range, where phonon-induced charge displacement is maximized. At the ballistic phonon frequency near 125\,GHz, the induced charge decreases by approximately an order of magnitude but remains above \( 10^{-3}\,e \), well within the detection threshold of modern radio-frequency QPC sensors.
}
    \label{fig:induced_charge}
\end{figure}

Figure~\ref{fig:induced_charge} presents the calculated induced charge \( Q_{\text{ind}} \) on the QPC electrode as a function of phonon frequency, assuming a quantum dot--QPC separation of 1~\(\mu\)m. In the resonance region—where phonon frequencies fall between 10 and 30\,GHz and coincide with the natural oscillation frequencies of dipole-bound states—the induced charge reaches values up to \( 0.01\,e \). This represents a strong, readily measurable signal. However, this same frequency band is also partially populated by thermal phonons at cryogenic temperatures, introducing a significant background that can obscure signals from rare, energy-depositing events. Therefore, distinguishing thermally generated phonons from primary signal phonons becomes a critical challenge in maintaining high signal fidelity.

Outside the resonance band, the induced charge decreases significantly—by a factor of approximately 2 to 4 for holes and electrons—at the ballistic phonon frequency of approximately 125\,GHz, primarily due to reduced phonon-induced charge displacement efficiency. Despite this reduction, the induced signal remains above \( 10^{-3}\,e \) for holes and \( 10^{-2}\,e \) for electrons near this frequency. These charge levels are still well within the sensitivity range of modern radio-frequency quantum point contact (RF-QPC) sensors~\cite{Field1993, GonzalezZalba2015}, thereby reinforcing the feasibility of GeQuLEP’s phonon-based readout architecture for detecting ultra-low energy depositions with high fidelity.

As shown in Figures~\ref{fig:charge_displacement} and~\ref{fig:induced_charge}, the 10--30\,GHz frequency range is particularly effective for inducing measurable charge displacements via the deformation potential mechanism, thereby enhancing the signal response at the QPC. However, this frequency band is also partially populated by thermal phonons, which are significantly more abundant than the non-equilibrium phonons produced by rare processes such as dark matter scattering or neutrino interactions. At a cryogenic temperature of 4\,K, thermal phonons are primarily concentrated at frequencies below approximately 83\,GHz, but they can still contribute to the 10--30\,GHz range either directly through their thermal distribution or indirectly via anharmonic decay from higher-frequency modes. This spectral overlap complicates the interpretation of signals in this band, particularly in low-threshold detection regimes. Therefore, distinguishing thermally generated phonons from event-induced phonons is critical for improving the signal-to-noise ratio. To address this challenge, we develop a multifaceted strategy that exploits the inherent differences in the temporal behavior, spatial propagation, and spectral characteristics of thermal and signal phonons at 4\,K.

\subsection{Distinguishing Thermal and Primary Phonons at 4\,K}

At cryogenic temperatures, the thermal phonon population in a Ge crystal follows the Bose-Einstein distribution, with the occupation number at a given frequency \( f \) described by:
\[
n(f, T) = \frac{1}{\exp(hf / k_BT) - 1},
\]
where \( h \) is Planck’s constant, \( k_B \) is Boltzmann’s constant, and \( T \) is the absolute temperature. At 4\,K, the characteristic thermal phonon frequency is approximately \( f_{\text{thermal}} = k_B T / h \approx 83\,\mathrm{GHz} \). Although the thermal population in the 10--30\,GHz range is lower, it is still non-negligible, with occupation numbers of \( n \approx 2.1 \) at 10\,GHz and \( n \approx 0.3 \) at 30\,GHz~\cite{holland1963}. This overlap necessitates discrimination methods that go beyond frequency alone.

Signal phonons, on the contrary, are generated by localized energy deposition events such as CE$\nu$NS and dark-matter scattering. These events typically produce a primary phonon in the terahertz range, such as 1.8\,THz, which rapidly undergoes anharmonic decay. After approximately six generations of binary decay, the resulting daughter phonons populate the 10--30\,GHz band, yielding a burst of \( 2^6 = 64 \) phonons around 28\,GHz. These non-equilibrium phonons arrive in tightly correlated, sub-nanosecond bursts, in stark contrast to the uncorrelated, stochastic thermal phonon background.

This temporal clustering can be exploited using fast-timing phonon detectors such as superconducting transition-edge sensors (TESs) or QPCs proposed in this work~\cite{brink2006, ji2008}. For example, a burst of 64 phonons arriving within a 1\,ns window produces a statistically significant signal above the thermal background, which yields fewer than one phonon per nanosecond in the same frequency range. Additionally, the anisotropic phonon dispersion in Ge causes ballistic phonons to preferentially propagate along crystallographic axes~\cite{knoll}, enabling spatial discrimination using multi-channel phonon imaging arrays to reconstruct the event location and propagation direction.

\begin{figure} 
    \centering
    \includegraphics[width=0.85\linewidth]{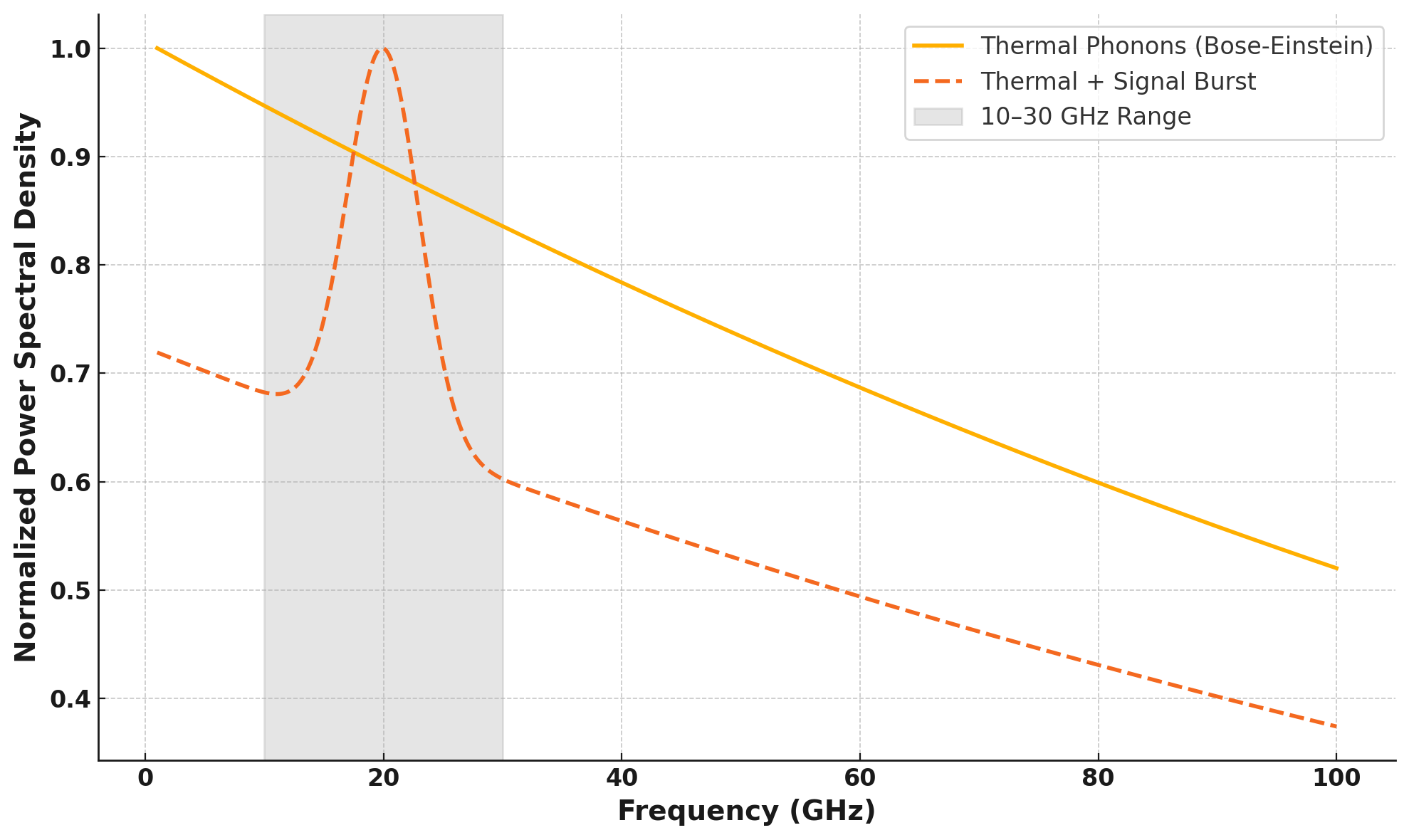}
    \caption{Normalized power spectral density of phonons in Ge at 4\,K. The solid line shows the thermal phonon background based on the Bose-Einstein distribution, while the dashed line includes an added Gaussian burst centered at 20\,GHz with a width of 3\,GHz, representing daughter phonons from the anharmonic decay of a 1.8\,THz primary phonon. The shaded region highlights the 10–30\,GHz band where discrimination is most effective. Both curves are normalized to a maximum value of 1.}
    \label{fig:phonon_psd}
\end{figure}

Figure~\ref{fig:phonon_psd} illustrates the normalized power spectral density (PSD) of phonons in Ge at 4\,K, comparing the thermal background spectrum with a simulated burst spectrum from an anharmonic decay. The thermal PSD is computed from the product \( f \cdot n(f, T) \), where \( n(f, T) \) follows the Bose-Einstein distribution. The burst component is modeled as a Gaussian distribution centered at 20\,GHz with a 3\,GHz bandwidth, consistent with the expected daughter phonon energies. Although the burst is spectrally embedded within the thermal background, it produces a localized enhancement in the PSD, particularly in the 10--30\,GHz range. This contrast provides a spectral window in which rare-event-induced phonons can be separated from the thermal population, especially when combined with time- and direction-sensitive detection.

To further suppress thermal phonons while enhancing the signal response, we incorporate spectral filtering using PnC cavities  engineered to exhibit bandgaps in the 100--120\,GHz range. A periodic array of sub-wavelength air holes in the QW region (e.g., with lattice constant \( a \approx 25\,\mathrm{nm} \)) creates a Bragg-reflection bandgap, satisfying the condition \( f \approx v_s / (2a) \), where \( v_s \approx 5400\,\mathrm{m/s} \) is the average speed of sound in Ge~\cite{maldovan2013}. Phonons resonant with the cavity’s defect modes are trapped and enhanced, while out-of-band thermal phonons are suppressed via reflection or scattering. This filtering capability, when combined with timing and spatial analysis, enables robust discrimination of signal phonons even in the presence of thermal background.

\section{Experimental Feasibility and Implementation Pathway}
\label{sec:experiment_outlook}

Although the theoretical foundation of the GeQuLEP detector architecture demonstrates strong potential for phonon spectroscopy, the realization of a practical, working prototype requires addressing several experimental challenges. Each subsystem—from phonon confinement and quantum dot stability to charge transduction and RF readout—must be precisely engineered and validated under cryogenic conditions.

\textbf{Stability and Reproducibility of Dipole-Induced Quantum Dots:} 
The GeQuLEP platform relies on quantum dots formed by the freeze-out of shallow impurities in high-purity Ge at temperatures below 10\,K. These dipole-defined quantum dots must maintain reproducible energy levels and charge stability to ensure consistent charge–phonon coupling. Achieving this requires precise control of impurity concentration and spatial distribution through controlled doping, crystal growth, and post-growth thermal annealing. Validation techniques such as low-temperature electron tunneling spectroscopy and RF reflectometry are essential to confirm quantum dot reproducibility, occupancy, and coupling consistency.

\textbf{Fabrication and Integration of PnC Structures on Ge:}  
Phonon confinement, directional control, and thermal noise suppression in the GeQuLEP architecture are enabled through the integration of high-precision PnC structures. This PnC cavity—functioning as both waveguides and spectral filters—is engineered to exhibit acoustic bandgaps in the 100--120\,GHz range along specific crystallographic axes. This bandgap effectively suppresses thermal phonons below $\sim$80\,GHz, which dominate the background at 4\,K, while permitting the transmission of ballistic, non-equilibrium phonons toward the sensing region.

To achieve the required spectral selectivity and phonon-guiding performance, sub-wavelength features with lateral dimensions around 25\,nm must be patterned with high fidelity and structural integrity. This poses significant fabrication challenges due to the mechanical softness and anisotropic etch behavior of Ge. These challenges are addressed through a combination of high-resolution lithographic and etching techniques, including electron-beam lithography (EBL), deep-ultraviolet (DUV) lithography, and reactive ion etching (RIE), employing robust hard masks such as silicon nitride (SiN) or chromium (Cr)~\cite{Zhu2017, Bahari2019}. To further enhance precision and compensate for fabrication-induced deviations, post-patterning tuning using focused ion beam (FIB) milling may be applied to locally modify unit cell dimensions or correct defects, thereby refining the phononic bandgap response.

The phononic filter itself comprises a periodic array of sub-wavelength elements—such as etched holes, nanopillars, or acoustic resonators—monolithically integrated onto the surface of a high-purity Ge crystal. It is typically situated between the phonon generation region and the dipole-defined quantum dot array. Despite its spectral filtering role, the PnC layer remains only tens to hundreds of nanometers thick, preserving compatibility with planar device geometries and ensuring minimal disruption to the detector’s mechanical or electrical configuration.

Importantly, the phononic filter is fully compatible with CMOS-friendly fabrication and layout practices, supporting scalable integration of QPCs dots. By selectively blocking incoherent thermal phonons while allowing signal-carrying ballistic phonons to propagate, the PnC significantly enhances the signal-to-noise ratio in the target detection band. This functionality is essential for achieving single-phonon resolution and enabling the GeQuLEP platform to detect rare events such as low-mass dark matter interactions and solar \( pp \)-neutrinos through coherent elastic neutrino–nucleus scattering (CE$\nu$NS).

\textbf{QPC Sensitivity and Cryogenic Noise Mitigation:}
Resolving single-phonon-induced charge displacements requires ultra-sensitive QPC readout under 4~K temperatures. Potential challenges include gate leakage, charge noise, and interface defects. These can be mitigated through atomic-layer deposition (ALD) of AlOx for surface passivation, and by integrating superconducting or RF tank circuits with the QPC for low-noise reflectometry readout~\cite{GonzalezZalba2015, Chorley2012, Cassidy2007, Colless2013}. These approaches have proven effective in GaAs- and Si/SiGe-based systems and are expected to be adaptable to the Ge platform.

\textbf{Phonon Lifetime and Collection Efficiency:}
The successful operation of the GeQuLEP sensor hinges on ballistic phonon survival over centimeter-scale distances. Theoretical predictions and experimental results from cryogenic detectors such as SuperCDMS and EDELWEISS suggest that LA phonons in high-purity Ge have lifetimes on the order of 1\,$\mu$s at temperatures below 1\,K~\cite{Agnese2014, Armengaud2017}. However, imperfections such as surface roughness and thin native oxide layers can introduce diffuse scattering and phonon decoherence. Mitigation strategies include chemical mechanical polishing (CMP), hydrogen annealing, and in-situ passivation, all of which are crucial for preserving phonon coherence and maximizing charge–phonon interaction fidelity.

\textbf{Spatial Alignment of Quantum Dots and QPC Sensors:}
Efficient transduction of phonon-induced lattice displacements requires precise alignment between the dipole-induced quantum dots and adjacent QPC readouts. Sputter doping with post-annealing can localize shallow impurity sites with sub-100\,nm accuracy. QPC geometries patterned by lithography can then be aligned relative to these sites, achieving lateral separations of 50–200\,nm—sufficient for optimal capacitive coupling while maintaining low parasitic noise. These configurations are Complementary Metal–Oxide–Semiconductor (CMOS)-compatible and scalable, enabling systematic integration of quantum sensors.

\textbf{Prototyping Pathway:}
A viable experimental pathway for constructing the first GeQuLEP prototype---compact in size with dimensions of 2\,cm in diameter and 1\,cm in thickness (approximately 17\,g)---involves bonding or epitaxially growing a thin Ge layer onto an insulating substrate. Gate-defined QPCs and PnC patterns can be fabricated via EBL or DUV lithography. SAW transducers may be placed at device boundaries to verify phonon transport and confinement. Phonon-induced signals can be detected by modulating phonon input and recording corresponding shifts in QPC conductance using RF reflectometry. These experimental strategies are well-established in quantum device research and are readily transferable to GeQuLEP development.

Overall, while challenging, the outlined fabrication and characterization steps fall within current state-of-the-art capabilities. The successful realization of GeQuLEP will require interdisciplinary collaboration across quantum electronics, cryogenic physics, semiconductor nanofabrication, and materials science. The feasibility of this architecture—backed by precedent in high-purity Ge detectors and RF-QPC systems—positions GeQuLEP as a compelling next-generation phonon spectroscopy for phonon-based quantum sensing.

\subsection{Discussion}

\begin{figure} [H]
    \centering
    \includegraphics[width=0.75\textwidth]{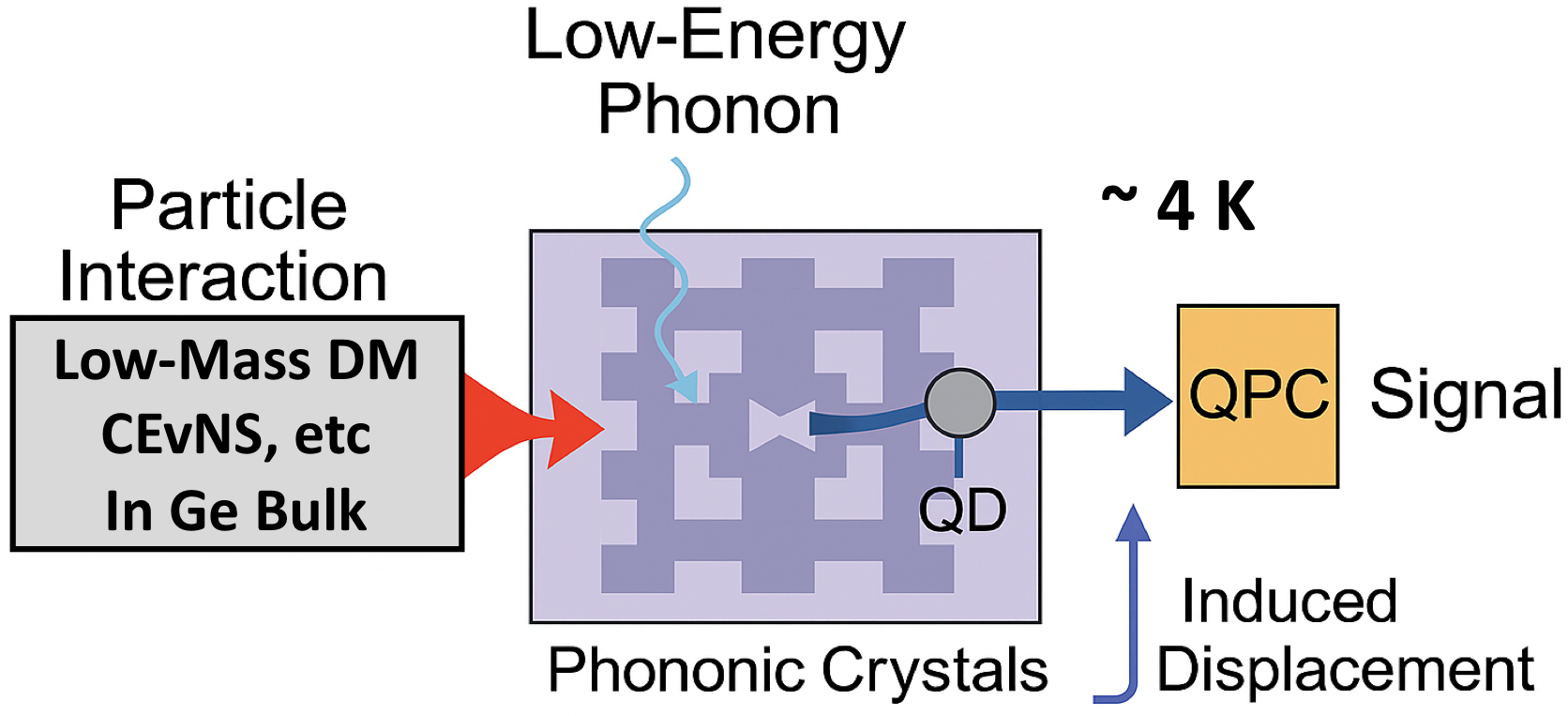}
    \caption{Conceptual block diagram of the GeQuLEP detection system. Energy deposited by low-energy particles in the Ge bulk generates primary phonons, which undergo anharmonic decay into ballistic acoustic phonons. These phonons are filtered and directed by a thin PnC cavity layer that defines the QW region. Within this region, the phonons interact with dipole-localized charge carriers, inducing measurable displacements. These displacements are detected by a nearby RF-QPC readout system. The overall architecture enables sub-eV energy sensitivity, making it suitable for rare-event detection such as low-mass dark matter and neutrino interactions.
}
    \label{fig:block_diagram}
\end{figure}

To synthesize the key elements of the GeQuLEP detection architecture, Figure~\ref{fig:block_diagram} provides a conceptual block diagram illustrating the full signal transduction chain—from phonon generation to charge readout. Energy depositions from rare interactions, such as dark matter scattering or CE$\nu$NS, produce athermal phonons that propagate ballistically through the high-purity Ge detector volume. These phonons are selectively guided by surface PnC circuits toward localized regions containing dipole-bound quantum dots. Phonon-induced lattice and charge displacements modulate the position of bound charges via the deformation potential, which in turn induces a measurable electrostatic signal. This signal is transduced by a nearby QPC, yielding a conductance shift proportional to the phonon-induced charge displacement. The entire system is operated at cryogenic temperatures (\(\sim 4~\text{K}\)) to suppress thermal phonon noise and preserve coherence. This diagram conceptually integrates the phononic, electronic, and quantum sensing elements discussed throughout the manuscript.

To evaluate the potential of GeQuLEP for detecting low-energy signals from dark matter and neutrino interactions, we compare its projected performance with existing leading experiments, including XENONnT~\cite{Aprile2022}, LUX-ZEPLIN (LZ)~\cite{Akerib2023}, SuperCDMS~\cite{Agnese2014}, and EDELWEISS~\cite{Armengaud2017}. The primary advantage of GeQuLEP lies in its ability to detect energy depositions down to the single-phonon level (\(\sim 0.00745~\text{eV}\)), offering several order-of-magnitude improvement in threshold sensitivity and enabling detection of low-mass dark matter and solar \(pp\)-neutrinos.

Specifically, the GeQuLEP detector is expected to provide:
\begin{enumerate}
    \item \textbf{Ultra-low Energy Threshold:} A projected threshold near \(0.00745~\text{eV}\) allows GeQuLEP to access previously unexplored kinematic regimes for low-mass dark matter and CE$\nu$NS.
    \item \textbf{High Detection Efficiency:} Optimized phonon guiding and near-unity phonon-to-charge transduction enhance efficiency beyond that of conventional bolometric or liquid noble detectors.
    \item \textbf{Scalability:} Fabricated using CMOS-compatible processes, GeQuLEP supports scalable array integration, enabling deployment in large-mass rare-event experiments.
\end{enumerate}

A comparison of key performance metrics is presented in Table~\ref{tab:comparison}.

\begin{table}[ht]
    \centering
    \caption{Comparison of key metrics between GeQuLEP and existing leading low-energy detector technologies. This work$^{*}$ is a theoretical prediction.}
    \label{tab:comparison}
    \renewcommand{\arraystretch}{1.3}
    \begin{tabular}{lcccc}
    \hline\hline
    Detector & Threshold & Efficiency & Scalability & Temperature \\
    \hline
    XENONnT~\cite{Aprile2022} & \(\sim 1\,\text{keV}\) & High & High & 170~K \\
    LUX-ZEPLIN (LZ)~\cite{Akerib2023} & \(\sim 1\,\text{keV}\) & High & High & 170~K \\
    SuperCDMS~\cite{Agnese2014} & \(\sim 56\,\text{eV}\) & Medium & Moderate & 40~mK \\
    EDELWEISS~\cite{Armengaud2017} & \(\sim 700\,\text{eV}\) & Medium & Moderate & 10~mK \\
    \textbf{GeQuLEP (this work$^{*}$)} & \(\sim 0.00745\,\text{eV}\) & Medium & Moderate & 4~K \\
    \hline\hline
    \end{tabular}
\end{table}

Further comparison with ongoing experimental efforts targeting MeV-scale dark matter highlights the unique promise of GeQuLEP as a next-generation phonon spectroscopy platform. Existing detectors—such as SuperCDMS~\cite{Agnese2018}, EDELWEISS~\cite{Armengaud2019}, SENSEI~\cite{Barak2020}, DAMIC~\cite{AguilarArevalo2019}, and NEWS-G~\cite{Arnaud2018}—have achieved impressive energy thresholds in the eV range. However, they remain fundamentally constrained by limitations in phonon and charge sensor resolution, particularly at ultra-low energies. By contrast, the GeQuLEP architecture is explicitly engineered to function as a high-resolution phonon spectrometer capable of resolving single primary phonon events with energy thresholds as low as \( 0.00745~\text{eV} \). This capability opens access to the keV/\(c^2\)-scale dark matter mass range, a region that remains largely uncharted in current direct detection searches. As a phonon spectroscopy platform, GeQuLEP not only pushes the sensitivity frontier but also introduces a new paradigm in rare-event detection through quantum-enhanced phonon readout.

The sensitivity of a rare-event detector like GeQuLEP—expressed in terms of the dark matter–nucleus scattering cross section as a function of particle mass—is not solely determined by its energy threshold. While the sub-eV sensitivity of GeQuLEP significantly enhances the reach for detecting low-mass dark matter particles, the ultimate detector performance is strongly influenced by the surrounding noise environment and background events. These include thermal phonons, charge fluctuations, and electromagnetic interference, all of which can obscure or mimic true signal events, especially at cryogenic operating temperatures. To produce reliable sensitivity curves and robust exclusion limits, these background contributions must be thoroughly characterized and effectively suppressed. Consequently, the construction and testing of a functional prototype is indispensable—not only to demonstrate the feasibility of phonon-to-charge transduction via RF-QPC readout, but also to establish credible background levels and validate the projected detection limits for dark matter and other low-energy interactions.

In addition to its dark matter detection capabilities, the GeQuLEP architecture holds considerable promise for observing CE$\nu$NS from solar \(pp\)-neutrinos. Preliminary calculations suggest that an exposure as small as \(10\text{--}100\,\text{g-day}\) could yield a statistically significant event rate, provided that background levels remain comparable to those achieved in current underground cryogenic detectors. This highlights the scientific versatility of the GeQuLEP platform, extending its relevance to neutrino physics and low-energy astrophysical phenomena.

Operating such a sensitive detector at cryogenic temperatures requires the use of dilution refrigerators, whose mechanical compressors unfortunately introduce low-frequency vibrations, typically in the 1--300\,Hz range. These classical oscillations, with macroscopic wavelengths and $\mu$eV-scale energies, differ fundamentally from the high-frequency (10--100\,GHz) acoustic phonons that effectively couple to dipole-localized charge carriers in Ge. At 4\,K, the confined carriers respond primarily to quantized lattice vibrations in the meV range, where the deformation-potential interaction is strongest. As a result, the long-wavelength compressor-induced vibrations are unlikely to directly excite dipole states or generate meaningful charge displacement. Nonetheless, secondary effects—such as microphonic pickup in the RF-QPC circuitry or stress-induced perturbations in device geometry—can still introduce spurious noise. Although these effects do not threaten the phonon transduction mechanism directly, they must be addressed through robust vibration isolation strategies to ensure low-noise operation.

At 4\,K, the charge sensitivity of QPC-based readout systems is largely dictated by thermal, shot, and charge fluctuation noise. Experimental results show that RF-QPCs operated under cryogenic conditions can reach charge sensitivities on the order of \( \delta q \sim 10^{-3}e/\sqrt{\mathrm{Hz}} \), and in optimized systems, down to \( \sim 10^{-4}e/\sqrt{\mathrm{Hz}} \) within MHz bandwidths~\cite{Cassidy2007, Reilly2007}. These sensitivities translate to an equivalent noise power near \( 10^{-18}~\mathrm{W/\sqrt{Hz}} \), positioning RF-QPCs as ideal transducers for detecting the tiny electrostatic variations induced by single- or few-phonon events. This makes them a key enabler of the GeQuLEP architecture's ability to probe rare sub-eV phenomena with exceptional precision.

If a ballistic phonon (from a primary phonon) interaction induces a charge displacement on the order of \(10^{-3}e\), the resulting signal remains above the noise floor of standard QPCs operating at 4\,K within a 1\,Hz bandwidth. While this signal magnitude is close to the lower detection limit, it becomes measurable under optimized conditions—particularly in narrow-band or time-integrated detection schemes where the signal-to-noise ratio (SNR) can be boosted above unity. Additionally, the integration of impedance-matched RF circuits with cryogenic amplification enables RF-QPCs to achieve charge sensitivity at or below \(10^{-3}e\), as demonstrated in previous studies~\cite{Cassidy2007}. These advancements underscore the feasibility of resolving such small charge displacements in the GeQuLEP architecture, thereby supporting its role as a viable platform for phonon spectroscopy and ultra-sensitive quantum sensing.

Prototype development is currently underway, with a focus on fabricating and integrating dipole-defined quantum dots and QPC readout structures within high-purity Ge substrates. The immediate experimental goal is to demonstrate phonon detection efficiency by measuring charge displacement induced by controlled phonon input using RF-QPC sensing. This includes benchmarking sensitivity and noise performance at cryogenic temperatures, as well as validating phonon confinement and guidance through lithographically patterned PnC cavities. These near-term milestones are critical for demonstrating proof of concept and advancing GeQuLEP toward deployment in next-generation quantum sensors for low-energy physics.

\section{Conclusion}
\label{sec:conclusion}

We have introduced \textbf{GeQuLEP} as a novel platform for \textit{phonon spectroscopy} and quantum sensing, offering ultra-sensitive detection of sub-eV energy depositions. By integrating high-purity Ge crystals with dipole-defined quantum dots and PnC cavities, GeQuLEP utilizes deformation-potential coupling to transduce energy from athermal phonons into measurable charge signals via RF-QPCs. Unlike conventional detector technologies that rely on bulk charge transport and electrical contacts, GeQuLEP harnesses ballistic phonons as quantum messengers—opening a new paradigm in rare-event detection and condensed-matter spectroscopy.

Our theoretical modeling demonstrates that phonons generated from low-energy processes such as dark matter scattering and CE$\nu$NS can induce localized charge displacements on the order of nanometers. These displacements result in detectable signals on adjacent QPC electrodes, even at frequencies above 100\,GHz. Resonant enhancement in the 10--30\,GHz range—where phonon frequencies match the natural oscillation modes of dipole-bound charges—further boosts the induced charge to values near \(0.01\,e\). While this regime overlaps with the thermal phonon background at 4\,K, GeQuLEP mitigates such noise through two key strategies: (1) spectral filtering using PnC cavities to suppress incoherent thermal phonons, and (2) time-domain discrimination leveraging the burst profile of ballistic primary phonons. These methods enhance the platform's capability to distinguish signal from background, reinforcing its functionality as a high-resolution phonon spectrometer.

The design further incorporates embedded PnC cavities within doped dipole layers and integrates an ultra-thin frequency-selective phononic filter, allowing for precise phonon confinement and energy focusing near the QPC. This structural optimization supports both spatial and spectral control of phonon propagation, ensuring efficient charge--phonon transduction and minimal loss. Our model accounts for phonon generation, ballistic transport, surface reflections, and absorption mechanisms—all critical for preserving coherence and maximizing detection efficiency.

With a projected single-phonon sensitivity of \( \sim 0.00745~\text{eV} \), GeQuLEP stands as a transformative phonon spectroscopy platform for probing weakly interacting particles and rare energy deposition events. Its CMOS-compatible, planar architecture offers scalable deployment, while its contact-free sensing and absence of high-voltage requirements make it robust and low-noise. These features collectively position GeQuLEP at the forefront of next-generation quantum detectors—capable of exploring new regimes in low-mass dark matter searches, solar \(pp\)-neutrino detection, and fundamental studies in phonon-mediated quantum phenomena.

\section*{Acknowledgment}
This work was supported in part by NSF OISE 1743790, NSF PHYS 2310027, NSF OIA 2437416, DOE DE-SC0024519, DE-SC0004768 and a research center supported by the State of South Dakota.

\vspace{1em}
\noindent\textit{Correspondence:} Dongming Mei, University of South Dakota

\end{document}